%% file: SubsVsTolls_j4.tex
\documentclass[letterpaper,10 pt, journal]{IEEEtran}  

\IEEEoverridecommandlockouts                              

\usepackage[utf8]{inputenc}
\usepackage{include_packages} 
\usepackage[english]{babel}

\title{
The Effectiveness of Subsidies and Tolls in Congestion Games}
\author{Bryce L. Ferguson, Philip N. Brown, and Jason R. Marden
\thanks{This research was supported by ONR grant \#N00014-17-1-2060, AFOSR grant \#FA9550-20-1-0054, and NSF grant \#ECCS-2013779.}
\thanks{B. L. Ferguson and J. R. Marden are with the Department of Electrical and Computer Engineering, University of California, Santa Barbara, CA, {\texttt{\{blferguson,jrmarden\}@ece.ucsb.edu}}.}
\thanks{P. N. Brown is with the Department of Computer Science, University of Colorado at Colorado Springs, {\texttt{philip.brown@uccs.edu}}}
\thanks{The conference version of this paper appeared in~\cite{Ferguson2020c}}
}

\begin{document}
\maketitle

\begin{abstract}
Are rewards or penalties more effective in influencing user behavior?
This work compares the effectiveness of subsidies and tolls in incentivizing user behavior in congestion games.
The predominantly studied method of influencing user behavior in network routing problems is to institute taxes which alter users' observed costs in a manner that causes their self-interested choices to more closely align with a system-level objective.
Another conceivable method to accomplish the same goal is to subsidize the users' actions that are preferable from a system-level perspective.
We show that, when users behave similarly and predictably, subsidies offer superior performance guarantees to tolls under similar budgetary constraints; however, in the presence of unknown player heterogeneity, subsidies fail to offer the same robustness as tolls.
\end{abstract}

\input{Introduction}

\input{Preliminaries}

\input{Bounded_Incentives}

\input{Incentives_w_Hetero}

\input{Robustness}
\input{Robust_and_bounds}
\input{Conclusion}

\bibliographystyle{IEEEtran}
\bibliography{../../../../../library}

\input{Appendix}

\begin{IEEEbiography}
    [{\includegraphics[width=1in,height=1.25in,clip,keepaspectratio]{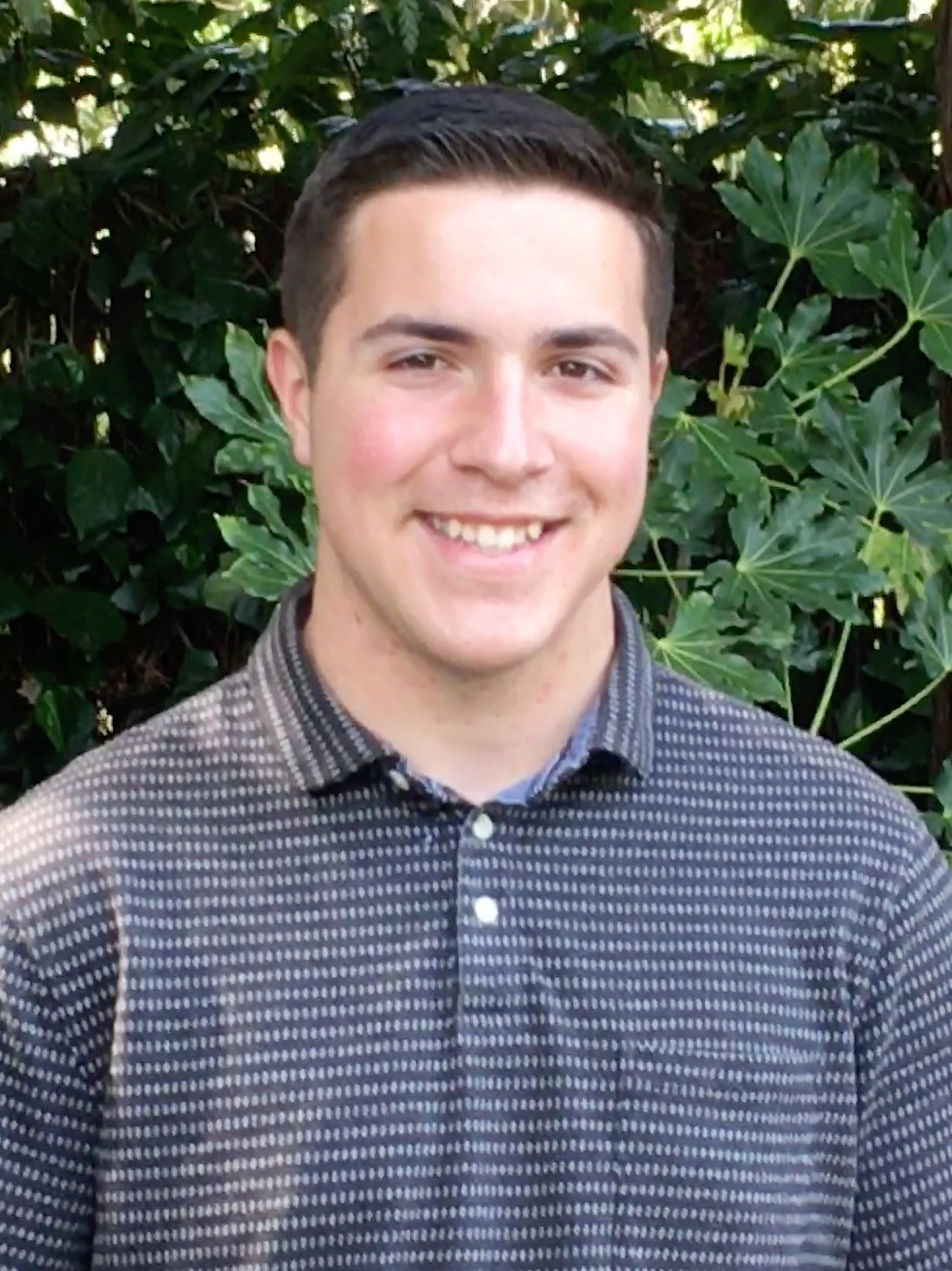}}]{Bryce L. Ferguson}
is a PhD student in the Electrical and Computer Engineering Department at the University of California, Santa Barbara. Bryce received his BS and MS in Electrical Engineering from the University of California, Santa Barbara in June 2018 and March 2020, respectively.
He is a finalist for the Best Student Paper Award at the 2020 IEEE American Controls Conference.
Bryce's research interests focus on using game theoretic methods for describing and controlling both societal and engineered multiagent systems.
\end{IEEEbiography}

\begin{IEEEbiography}
    [{\includegraphics[width=1in,height=1.25in,clip,keepaspectratio]{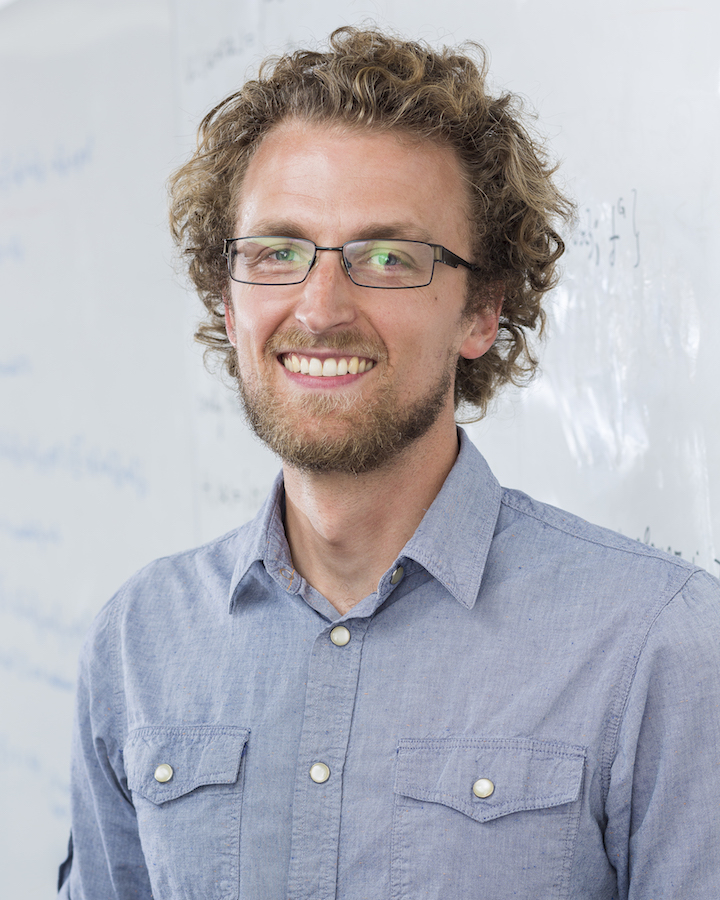}}]{Philip N. Brown}
is an Assistant Professor in the Department of Computer Science at the University of Colorado, Colorado Springs.
Philip received the Bachelor of Science in Electrical Engineering in 2007 from Georgia Tech, after which he spent several years designing control systems and process technology for the biodiesel industry.
He received the Master of Science in Electrical Engineering in 2015 from the University of Colorado at Boulder under the supervision of Jason R. Marden, where he was a recipient of the University of Colorado Chancellor's Fellowship.
He received the PhD in Electrical and Computer Engineering from the University of California, Santa Barbara under the supervision of Jason R. Marden.
He was finalist for the Best Student Paper Award at the 2016 and 2017 IEEE Conferences on Decision and Control, and received the 2018 CCDC Best PhD Thesis Award from UCSB.
Philip is interested in the interactions between engineered and social systems.
\end{IEEEbiography}

\begin{IEEEbiography}
    [{\includegraphics[width=1in,height=1.25in,clip,keepaspectratio]{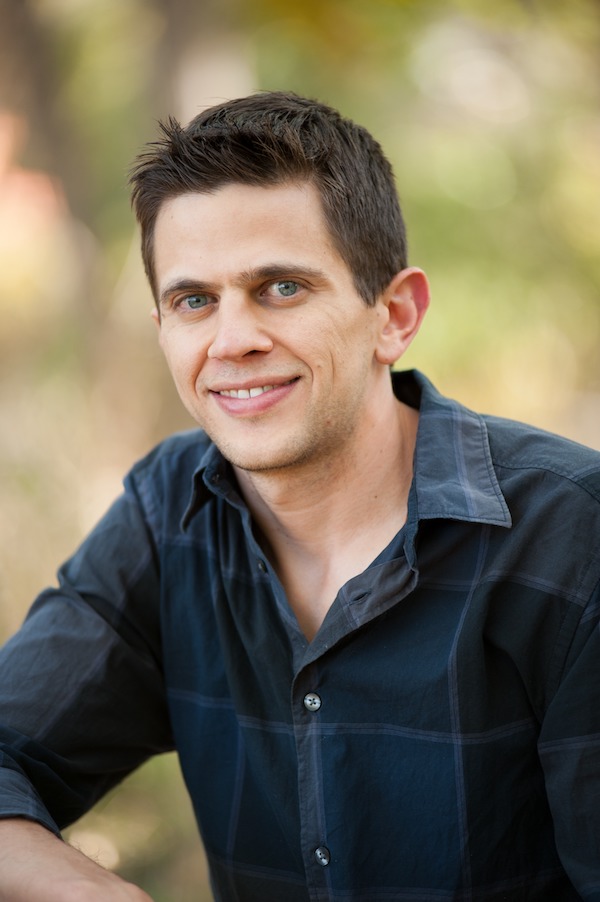}}]{Jason R. Marden}
is an Assistant Professor in the Department of Electrical and Computer Engineering at the University of California, Santa Barbara.
Jason received the Bachelor of Science in Mechanical Engineering in 2001 from UCLA, and the PhD in Mechanical Engineering in 2007, also from UCLA, under the supervision of Jeff S. Shamma, where he was awarded the Outstanding Graduating PhD Student in Mechanical Engineering.
After graduating from UCLA, he served as a junior fellow in the Social and Information Sciences Laboratory at the California Institute of Technology until 2010, and then as an Assistant Professor at the University of Colorado until 2015. Jason is a recipient of an ONR Young Investigator Award (2015), NSF Career Award (2014), AFOSR Young Investigator Award (2012), SIAM CST Best Sicon Paper Award (2015), and the American Automatic Control Council Donald P. Eckman Award (2012). Jason’s research interests focus on game theoretic methods for the control of distributed multiagent syste
\end{IEEEbiography}

\end{document}

%% file: Introduction.tex
\section{Introduction}
In systems governed by a collective of multiple decision making users, system performance is often dictated by the choices those users make.
Though each user may make decisions rationally, the emergent behavior observed in the system need not align with the objective of the system designer.
This phenomenon appears in many engineering settings including distributed control~\cite{shamma2007cooperative}, resource allocation problems~\cite{Johari2010}, electric power grids \cite{Palensky2011}, and transportation networks~\cite{salazar2019congestion}, as well as many logistical problem settings such as marketing~\cite{Narasimhan1984} and supply-chain management~\cite{Esmaeili2009}.
A prominent metric to quantify this emergent inefficiency is the \emph{price of anarchy}, defined as the worst-case ratio between the social welfare experienced when users make self interested decisions and the optimal social welfare~\cite{Papadimitriou2001,Kleer2017}.

A promising method of mitigating this inefficiency is by introducing incentives to the system's users, influencing their decisions to more closely align with the system optimal~\cite{Ratliff2018}.
One example of such incentives is to levy \emph{taxes}, eliciting monetary fees from users will affect their preferences over the available actions (e.g., tolls in transportation)~\cite{Ferguson2019,Cole2003,Fotakis2010}.
Such taxes have been shown to be effective in reducing system inefficiency as measured by the price of anarchy ratio~\cite{Bilo2016,Caragiannis2010,Paccagnan2019I,Fleischer2004}.
\BLF{Another method to influence user behavior is to \emph{subsidize} the actions that are preferable from a system level perspective.}
Subsidies have been studied as a tool to influence users in transportation~\cite{Zou2019}, supply chains~\cite{Taylor2002}, congestion~\cite{Bagolee2017}, and emissions~\cite{Chen2012emissions}.
Though subsidies require the system operator to pay its users, it is possible that the savings obtained from efficient use of the infrastructure outweigh the cost incurred from the implemented incentives~\cite{Gerstner1991,Ali1994}; additionally, one could consider implementing subsidies as rebates to a fixed, opt-in fee, to prevent a loss of revenue for the system operator.
Though the use of subsidies is feasible in theory and in implementation, this method has been studied significantly less than the tax equivalent; the relative performance of each is thus unknown.

In this paper, we seek to understand the relative performance of subsidies and taxes in influencing user behavior in socio-technical systems. 
Specifically, we consider a network routing problem in which users must traverse a network with congestible edges with delays that grow as a function of the local mass of users.
Finding a route for each user that minimizes the total latency in the system is straightforward if the system designer has full control in directing the users.
However, when users select their own routes, the resulting network flow need not be optimal~\cite{Pigou1920}.
Modeling the selfish routing problem as a congestion game, we adopt the \emph{Nash flow} as a solution concept of the emergent behavior in the system.
From the users' selfish routing, the price of anarchy may be large~\cite{Roughgarden2005}.
To alleviate this emergent inefficiency, we introduce incentives to the users' which alter their observed costs and preferences.
The objective of such incentives is to shape the users' preferences so the performance of the resulting Nash flow will improve.

A well studied method of incentivizing users in congestion games is to tax the users, i.e., introducing tolls to links in the network~\cite{Ferguson2019,Cole2003,Fotakis2010,Bilo2016,Caragiannis2010,Paccagnan2019I,Fleischer2004,Bonifaci2011,Fotakis2007}.
\BLF{In each of these referenced works, the price of anarchy is used to measure the effectiveness of a tolling scheme.}
Indeed in the most elementary settings, tolls exist that influence users to self route in line with the system optimum~\cite{Pigou1920}.
However, when more nuance is introduced in the form of \emph{player heterogeneity} (i.e., players differing in their response to incentives), the task of designing tolls becomes more involved.
When the toll designer possesses sufficient knowledge of the network structure and user population, they may still compute and implement tolls which incentivize optimal routing~\cite{Fleischer2004}.
However, in the case where the system designer has some uncertainty in the network parameters or behavior of the user population, it may not be possible to design tolls that give optimal system performance; thus, tolls are often designed to minimize inefficiency measured by the price of anarchy ratio~\cite{Fotakis2010,Karakostas2004,Brown2017d}, and again, encouraging results exist.

Though the study of tolling in congestion games is extensive, there are few results regarding subsidies as incentives in this context, especially in the presence of uncertain user heterogeneity.
In~\cite{Arieli2015}, the authors investigate budget-balanced tolls in which the sum of all monetary transactions is zero, but the authors only consider homogeneous users.
The authors of \cite{Maille2012} give the first formal analysis of subsidies in congestion games and provide an algorithm that computes optimal rebates when users are homogeneous and the network structure is known.
The authors of \cite{Sandholm2002} consider more general incentives, but in an evolutionary setting.
From a system designer's perspective, subsidies may be a feasible method of influencing user behavior; the performance guarantees of subsidies is thus of interest as well as how this performance compares to tax incentives.

Though there is a clear disparity in the breadth of results in the literature on tolls and subsidies, we bridge this gap by proving fundamental relationships between the performance and robustness of subsidies and tolls. Namely, subsidies offer better performance guarantees than tolls under budgetary constraints but are inherently less robust to user heterogeneity.
The manuscript is outlined as follows:

\begin{hangparas}{.25in}{1}
\textbf{\cref{sec:bounded}: Performance of Incentives.} \BLF{In \cref{thm:bounded}, it is shown in the nominal setting, where users behave similarly and predictably, that subsidies give better performance guarantees under similar budgetary constraints.}
\end{hangparas}
\begin{hangparas}{.25in}{1}
\textbf{\cref{sec:inc_w_hetero}: Incentives with Heterogeneity.} In \cref{thm:lim} it is shown that tolls can effectively mitigate the negative effects of player heterogeneity while in \cref{thm:lim_sub} it is shown subsidies cannot.
\end{hangparas}
\begin{hangparas}{.25in}{1}
\textbf{\cref{sec:robust}: Robustness of Incentives.} It is shown that tolls are more robust to uncertainty in the user population than subsidies.
In the presence of a budgetary constraint, \cref{thm:robust} shows that uncertainty degrades subsidy performance more rapidly than it degrades toll performance.
\end{hangparas}
\begin{hangparas}{.25in}{1}
\textbf{\cref{sec:bounded_and_robust}: Trade-off in Performance and Robustness.} Given the contrast in the nominal performance of subsidies and the robustness of tolls to user heterogeneity, this fundamental relationship is analyzed between the two in parallel-affine congestion games by finding the level of uncertainty at which the robustness of tolls gives superior performance guarantees than subsidies.
\end{hangparas}

In addition to finding general performance and robustness relationships between subsidies and tolls, we additionally find explicit price of anarchy bounds for optimal tolls and subsidies in several classes of congestion games to show that the differences in performance can be significant.
We introduce tools to construct optimal incentives and corresponding performance guarantees.

%% file: Preliminaries.tex

\section{Preliminaries}\label{sec:prelim}
\subsection{System Model}
Consider a directed graph $(V,E)$ with vertex set $V$, edge set $E \subseteq (V \times V)$, and $k$ origin-destination pairs $(o_i,d_i)$.
Denote by $\paths_i$ the set of all simple paths connecting origin $o_i$ to destination $d_i$.
Further, let $\paths = \cup_{i=1}^k \paths_i$ denote the set of all paths in the graph.
A \textit{flow} on the graph is a vector $f \in \mathbb{R}^{|\paths|}_{\geq 0}$ that expresses the mass of traffic utilizing each path.
The mass of traffic on an edge $e \in E$ is thus \BLF{$f_e = \sum_{P:e\in P} f_P$}, and we say $f = \{f_e\}_{e \in E}$.
A flow $f$ is \textit{feasible} if it satisfies $\sum_{P \in \paths_i}f_P = r_i$ for each source-destination pair, where $r_i$ is the mass of traffic traveling from origin $o_i$ to destination $d_i$.

\begin{figure}[t]
\centering
    \begin{tikzpicture}[thick,scale=1.2, every node/.style={scale=1.2}]
        \node () [align=center, below] at (2.5,2) {$e_1$};
        \node () [align=left, below] at (1.4,1.6) {$e_2$};
        \node () [align=center, above] at (1,0.5) {$e_3$};
        \node () [align=center, below] at (1.5,0) {$e_4$};
        \node () [align=center, below] at (2.4,0.6) {$e_5$};
        \node (v1) [circle,draw,inner sep=0pt,fill=white,minimum width=0.5cm]  at (1.0,2.0) {$v_1$};
        \node (v2) [circle,draw,inner sep=0pt,fill=white,minimum width=0.5cm]  at (0.0,0) {$v_2$};
        \node (v3) [circle,draw,inner sep=0pt,fill=white,minimum width=0.5cm]  at (2.0,1.0) {$v_3$};
        \node (v4) [circle,draw,inner sep=0pt,fill=white,minimum width=0.5cm]  at (3.0,0.0) {$v_4$};
        \draw [->,thick,out=0,in=180] (v2) edge (v4);
        \draw [->,thick] (v1) edge (v3);
        \draw [->,thick] (v2) edge (v3);
        \draw [->,thick] (v3) edge (v4);
        \draw [->,thick,out=0,in=90] (v1) edge (v4);
    \end{tikzpicture}
    ~~~~
		\begin{tabular}[b]{l|l}
		\hline
		\\[-1em]
		 $e_1$ & $\ell_1(f_1)=4f_1^2$  \\[-1em] \\ \hline \\[-1em]
		 $e_2$ & $\ell_2(f_2)=1/2$  \\[-1em] \\ \hline \\[-1em]
		 $e_3$ & $\ell_3(f_3)=1/2$  \\[-1em] \\ \hline \\[-1em]
		 $e_4$ & $\ell_4(f_4)=2f_4$  \\[-1em] \\ \hline \\[-1em]
		 $e_5$ & $\ell_5(f_5)=1/2$  \\[-1em] \\ \hline
		\end{tabular}
    \caption{\small An example network routing problem $G$ with two origin-destination pairs: $(o_1,d_1)=(v_1,v_4)$ with $r_1=1/2$, and $(o_2,d_2)=(v_2,v_4)$ with $r_2=1/2$.}
    \label{fig:routing_example}
\end{figure}
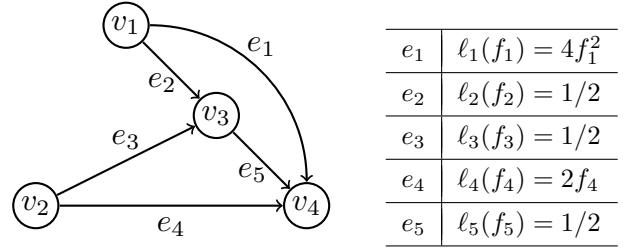

Each edge $e \in E$ in the network is endowed with a non-negative, non-decreasing \textit{latency} function $\ell_e:\mathbb{R}_{\geq 0} \rightarrow \mathbb{R}_{\geq 0}$ that maps the mass of traffic on an edge to the delay users on that edge observe.
The system cost of a flow $f$ is the \textit{total latency},
\begin{equation}\label{eq:total_lat}
	\mathcal{L}(f) = \sum_{e \in E} f_e \cdot \ell_e(f_e).
\end{equation}
A \textit{routing problem} is specified by the tuple $G = (V,E,\{\ell_e\}_{e\in E},\{r_i,(o_i,d_i)\}_{i=1}^k)$ as illustrated in \cref{fig:routing_example}, and we  let $\mathcal{F}(G)$ denote the set of all feasible flows.
We define the optimal flow $\fopt$ as one that minimizes the total latency, i.e.,
\begin{equation}
	\fopt \in \argmin_{f \in \mathcal{F}(G)} \mathcal{L}(f).
\end{equation}
We denote a family of routing problems by $\gee$.
\BLF{A family of routing problems is any set of routing problems, often specified by a specific network topology (e.g., parallel networks) and/or edge latency function types (e.g., polynomial latency functions) but can also be a singleton.}

\subsection{Incentives}
In this paper, we consider the problem of selfish routing, where each user in the system chooses a path as to minimize their own observed delay.
Let $N_i$ be the set of users traveling from origin $o_i$ to destination $d_i$.
\BLF{Each non-atomic user $x \in N_i$ is thus free to choose between paths $P \in \paths_i$.}
Let each $N_i$ be a closed interval with Lebesgue measure $\mu(N_i) = r_i$ that is disjoint from each other set of users, i.e., $N_i \cap N_j = \emptyset \ \forall \ i,j \in \{ 1,\ldots,k \}, \ i \neq j$.
The full set of agents is thus $N = \cup_{i=1}^k N_i$ whose mass is $\mu(N) = \sum_{i=1}^k r_i$.

It is well known that selfish routing can lead to sub-optimal system performance~\cite{Roughgarden2005}.
It is therefore up to a system designer to select a set of \textit{incentive functions} $\tau_e:\mathbb{R}_{\geq 0} \rightarrow \mathbb{R} \ \forall e\in E$ to influence the behavior of the users in the system to more closely align with the system optimal flow.
These incentives can be regarded as monetary transfers with the users dependent on the paths they choose.

A user $x \in N_i$ traveling on a path $P_x \in \paths_i$ observes cost
\begin{equation} \label{eq:player_cost}
	J_x(P_x,f) = \sum_{e \in P_x} \ell_e(f_e) + \tau_e(f_e).
\end{equation}
A flow $f$ is a \textit{Nash flow} if
\begin{multline}
	J_x(P_x,f) \in \argmin_{P \in \paths_i} \left\lbrace \sum_{e\in P} \ell_e(f_e) + \tau_e(f_e) \right\rbrace \\ \forall x\in N_i, ~ i \in \{1,\ldots,k\}.
\end{multline}

A game is therefore characterized by a routing problem $G$ and a set of incentive functions $\{\tau_e\}_{e\in E}$, denoted by the tuple $(G,\{\tau_e\}_{e\in E})$. It is shown in \cite{Mas-Colell1984} that a Nash flow exists in a congestion game of this form if the latency and incentive functions are Lebesgue-integrable.

\subsection{Incentive Mechanisms \& Performance Metrics}
To determine the manner in which incentive functions are applied to edges, we investigate \textit{incentive mechanisms}.
To formalize this notion, let
\BLF{$$L(G) := \{(\ell_e,e,G)\}_{e \in E(G)}$$ 
be the set of links or edges in the routing problem $G$.
An element in $L(G)$ is a tuple of the latency function $\ell_e$, edge index $e$, and routing problem $G$ for each link in the edge set $E(G)$.
Further, for a family of problems, we denote $L(\gee) = \cup_{G\in \gee}L(G)$ as the set of links that occur in the family of games $\gee$.

For each edge $e$ in the routing problem $G$ with latency function $\ell_e$, an incentive mechanism $T$ assigns an incentive $T(\ell_e;e,G)$, i.e. $\tau_e(f_e) = T(\ell_e;e,G)[f_e]$, where $T(\ell_e;e,G)[f_e]$ is the incentive evaluated at $f_e$.
This mapping is denoted by $T:L(\gee) \rightarrow \Tau$ where $\Tau$ is some set of allowable incentive functions.
For brevity, an incentive mechanism will be written simply as $T(\ell_e)$, but it is assumed that, unless otherwise stated, the incentive designer has knowledge of the exact edge and full network structure when assigning an incentive $T(\ell_e)$; these are termed \emph{network-aware} incentive mechanisms~\cite{Fleischer2004,Bilo2016}, and are the focus of \cref{thm:bounded} and \cref{thm:robust}.}

In the case where the incentive mechanism must be designed for a family of routing problems and without knowledge of the full network structure, we add the implied constraint that two edges with the same latency function are indistinguishable and must have the same assigned incentive;
we highlight such cases by terming the mechanism \emph{network-agnostic}.
\BLF{The use of network agnostic incentive mechanisms has been studied in \cite{Brown2017d,Paccagnan2019I} and are useful for their robustness in settings with frequent changes to the system structure (i.e., commerce, supply-chain-management, and even traffic when considering accidents and emergencies), where partial changes to the network structure or edge latencies need not require global redesign of the incentive mechanism.}
One such incentive that fits this framework is the classic Pigouvian or marginal cost tax,
\begin{equation} \label{eq:marginal_cost}
	T^{\rm mc}(\ell)[f] = f\cdot \frac{d}{df} \ell(f),
\end{equation}
which is known to incentivize users to route optimally in many classes of congestion games~\cite{Pigou1920}.
This is only true however, when there is no bound on the incentive and users are homogeneous~\cite{Brown2017d}.

We use the \textit{price of anarchy} to evaluate the performance of a taxation mechanism, defined as the worst case ratio between total latency in a Nash flow and an optimal flow, exemplified in~\cref{fig:routing_example}.
Let $\Lnash(G,T)$ be the highest total latency in a Nash flow of the game $(G,T(L(G)))$.
Additionally, let $\Lopt(G)$ be the total latency under the optimal flow $\fopt$.
The inefficiency can be characterized by
\vs
\begin{equation}
	\poa(G,T) = \frac{\Lnash(G,T)}{\Lopt(G)}.
\end{equation}
We extend this definition to a family of instances
\vs
\begin{equation}
	\poa(\gee,T) = \sup_{G \in \gee} \ \frac{\Lnash(G,T)}{\Lopt(G)},
\end{equation}
where $T$ is used in each routing problem.
The price of anarchy is now the worst case inefficiency over all such routing problems while using incentive mechanism $T$.
The objective of such incentive mechanisms is to minimize this worst case inefficiency, thus the optimal incentive mechanism is defined as,
\vs
\begin{equation}
	T^{\rm opt} \in \underset{T: L(\gee) \rightarrow \Tau}{\arginf} \poa(\gee,T),
\end{equation}
such that it minimizes the price of anarchy for a class of routing problems $\gee$.


\subsection{Tolls \& Subsidies}
We differentiate between two forms of incentives, \textit{tolls} $\tau_e^+:\mathbb{R}_{\geq 0} \rightarrow \mathbb{R}_{\geq 0}$ and \textit{subsidies} $\tau_e^-:\mathbb{R}_{\geq 0} \rightarrow \mathbb{R}_{\leq 0}$.
With tolls, the player's observed cost is strictly increased, i.e., the system designer levies taxes for the users to pay depending on their choice of edges.
With subsidies, the players cost is strictly reduced, i.e., the system designer offers some payments to users for their choice of action.
The main focus of this work is to assess which is more effective in influencing user behavior, tolls or subsidies.

A \textit{tolling mechanism} is one which only assigns tolling functions, defined as $T^+:L(\mathcal{G})\rightarrow \mathcal{T}^+$ where $\mathcal{T}^+$ is the set of \BLF{all non-negative, integrable functions on $\mathbb{R}^+$}.
An optimal tolling mechanism is one that minimizes the price of anarchy ratio, i.e.,
\begin{equation}
	T^{\rm opt+} \in \underset{T: L(\gee) \rightarrow \Tau^+}{\arginf} \poa(\gee,T).
\end{equation}
An optimal \textit{subsidy mechanism} is defined analogously with non-positive subsidy functions.
In the following sections, we compare the price of anarchy ratio associated with the optimal toll and optimal subsidy.

\BLF{The following example, illustrated in \cref{fig:routing_example}, highlights the notation and  the difference between tolls and subsidies.
\begin{example}\label{ex:homo}
Consider the network $G$ in \cref{fig:routing_example} with two origin destination pairs: $(o_1,d_1)=(v_1,v_4)$ with $r_1=1/2$, and $(o_2,d_2)=(v_2,v_4)$ with $r_2=1/2$.
The optimal flow in $G$, that minimizes \eqref{eq:total_lat}, is $\fopt \approx\{ 0.289,0.211,0.25,0.25,0.461 \}$ with a total latency of $\mathcal{L}(\fopt)\approx 0.683$.
With no tolling, the Nash flow is $\fnash=\{ 1/2,0,0,1/2,0 \}$ with total latency $\mathcal{L}(\fnash)=1$ producing a price of anarchy of $\poa(G,\emptyset)\approx 1.465$.
Under a scaled marginal-cost toll, the cost incurred by a user for utilizing edge $e$ is $\ell_e(f_e) + f_e\cdot \frac{d}{df_e}\ell_e(f_e)$ and the Nash flow becomes the same as $\fopt$, leading to a price of anarchy of $\poa(G,T^{\rm mc}) = 1$.
Similarly, under a subsidy mechanism $T^-(\ell_e) = \frac{1}{3}f_e\cdot \frac{d}{df_e}\ell_e(f_e) - \frac{2}{3}\ell_e$, the Nash flow is again the optimal, and $\poa(G,T^-) = 1$.
\end{example}
}

\BLF{This example highlights that subsidies and tolls are both effective at reducing the inefficiencies associated with selfish routing.
In this work, we study how this performance changes under budgetary constraints and user price-heterogeneity.}

\subsection{Summary of Our Contributions}
We start by addressing the nominal homogeneous setting, in which all users react to incentives identically.
In \cref{thm:bounded}, in any congestion game, we show that under a similar budgetary constraint, the optimal subsidy offers better performance than the optimal toll; the magnitude of this difference is exemplified in Proposition~\ref{prop:aff_bounded} by deriving explicit price of anarchy bounds for optimal tolls and subsidies in affine congestion games.

Next, we look at the efficacy of each incentive in mitigating the effect of user heterogeneity as the budgetary constraint is lifted.
In \cref{thm:lim}, we show that tolls can effectively eliminate the effect of user heterogeneity when the bound on incentives is lifted.
However, in \cref{thm:lim_sub}, it is shown that even in congestion games with convex, non-decreasing, continuously-differentiable latency functions, it is impossible for subsidies to mitigate the effect of user heterogeneity, even with the ability to give arbitrarily large payments.

When budgetary constraints do exist and users are heterogeneous in their response to incentives, we show in \cref{thm:robust} that, for tolls and subsidies bounded to give similar performance in the homogeneous setting, the performance of subsides is worse than tolls when users become heterogeneous, i.e., the performance of subsidies degrades more significantly from player heterogeneity than tolls. This is exemplified in Proposition~\ref{prop:aff_robust}, giving price of anarchy bounds for robust incentives in affine congestion games.

Finally, because subsides offer better performance under similar budgetary constraints in the homogeneous setting and tolls offer better robustness in the face of user heterogeneity, we investigate what level of user heterogeneity allows tolls to outperform similarly bounded subsides.
In \cref{thm:robust_and_bounded}, a relationship between the incentive bound and level of heterogeneity is derived in a class of parrallel-affine congestion games that leads to similar performance guarantees of the optimal toll and subsidy.



%% file: Bounded_Incentives.tex
\section{Bounded Incentives}\label{sec:bounded}
We first look at the case where users are homogeneous in their response to incentives.
This setting has been the focus of study for many incentive related works~\cite{Pigou1920,Paccagnan2019I,Bilo2016,Bonifaci2011,Fotakis2007}.
For these reasons, we start by comparing the effectiveness of subsidies and tolls in this setting when additional budgetary constraints are added. 
\BLF{Subsidies and tolls both serve as mechanisms for influencing user behavior and can be implemented by similar methods.
The act of applying constraints on either is of little difference to the system designer, however the reasoning for these constraints may differ. For instance, budgetary constraints on subsidies can serve to limit the monetary obligation of the system operator, while bounding tolls can prevent scenarios where users may avoid using the network entirely.
Though any specific budgetary constraint on either incentive is heavily influenced by the problem setting, here we seek to understand more generally how limits on the magnitude of incentives comparatively affect subsidies and tolls.}

To explore this, we introduce bounded tolls and subsidies.
A bounded toll satisfies $\tau_e^+(f_e) \in [0,\beta\cdot \ell_e(f_e)]$ for $f_e \geq 0$ and each $e \in E$, where $\beta$ is a bounding factor.
A bounded tolling mechanism is denoted by $T^+(\ell_e;\beta)$.
Similarly, a bounded subsidy satisfies $\tau_e^-(f_e) \in [-\beta\cdot \ell_e(f_e),0]$ for $f_e \geq 0$ and each $e \in E$, and a bounded subsidy mechanism is denoted by $T^-(\ell_e;\beta)$.
This form of bounded incentive functions resembles the bounded path deviations studied in \cite{Kleer2017}.
\BLF{Though many forms of bounding constraint can be considered, this form is chosen as it can be applied to network-aware and-network agnostic incentive mechanisms, captures the idea that larger delays can be incentivized more significantly, and avoids trivialities caused by arbitrarily large delays.
Additionally, these constraints can be represented as the total incentive in a routing problem being within a multiplicative factor $\beta$ of the total latency, i.e., $\sum_{e \in E} f_e |\tau_e(f_e)| \leq \beta \mathcal{L}(f)$.}

For some bounding factor $\beta$, let $\Tset_\beta^+$ denote the set of taxation mechanisms appropriately bounded by $\beta$.
More formally, $\Tset_\beta^+ = \{ T | T:L(\gee) \rightarrow \mathcal{T}^+(\beta) \}$, where 
\BLF{$$\mathcal{T}^+(\beta) = \{\tau_e^+ \in \Tau^+ \ | \ \tau_e^+(f_e) \in [0,\beta \cdot \ell_e(f_e)] ~ \forall f_e\geq 0 \}$$}
is the set of all tolling functions bounded by $\beta$.
To compare the efficacy of bounded tolls and subsidies, we define an optimal bounded tolling mechanism as 
\begin{equation}
	T^\mathrm{opt+}(\beta) \in \underset{T^+ \in \Tset_\beta^+}{\arginf} \ \poa(\gee,T^+).
\end{equation}
The optimal bounded subsidy mechanism $T^{\rm opt-}(\beta)$ is defined analogously.
\BLF{For brevity, bounded mechanisms are often written $T(\beta)$ when being discussed without reference to their use on a specific edge and $T(\ell_e;\beta)$ when they are referenced to a specific edge latency function.}

Though we consider any \BLF{incentive} bound $\beta \geq 0$, we offer the following definition to differentiate from cases where the bound is very large or trivially zero.

\begin{definition}\label{def:tight_bound}
A toll (subsidy) is \emph{tightly bounded} if \mbox{$\tau(f)=\beta\ell(f)$,} (if $\tau(f)=-\beta\ell(f)$) for some $f \geq 0$.
\end{definition}
When an optimal incentive is tightly bounded, the budgetary constraint is active.

\subsection{General Relation of Performance}
We first consider the relationship between bounded subsidies and tolls in general for congestion games (i.e., arbitrary latency functions and network topologies).
\cref{thm:bounded} states that bounded subsidies outperform similarly bounded tolls with respect to the price of anarchy, and strictly outperform when the budgetary constraint is active.

\begin{theorem}\label{thm:bounded}
For a congestion games $G$, under a bounding factor $\beta \geq 0$ the optimal subsidy mechanism $T^{\rm opt-}(\beta)$ has no greater price of anarchy than the optimal tolling mechanism $T^{\rm opt+}(\beta)$, i.e.,
\begin{equation}\label{eq:thmPoAincentive}
 	\poa \left(G,T^{\rm opt+}(\beta) \right) \geq \poa \left(G,T^{\rm opt-}(\beta) \right) \geq 1.
\end{equation}
Additionally, if every optimal subsidy is tightly bounded\footnote{\BLF{$T^{\rm opt-}(\ell_e;\beta)$ satisfies \cref{def:tight_bound} with bounding factor $\beta$ for each $\ell_e \in L(\gee)$.}}, then the first inequality in \eqref{eq:thmPoAincentive} is strict.
\end{theorem}
The proof of \cref{thm:bounded} appears at the end of this subsection; we first discuss the implications of this result.
\cref{thm:bounded} implies that when limiting the size of monetary transactions with homogeneous users, subsidies are more effective than tolls at influencing user behavior.
This result holds for any congestion game.
Though \eqref{eq:thmPoAincentive} need not be strict in general, there does exist a gap between the performance of tolls and subsidies in many non-trivial settings.
To illustrate this, we offer the following example to highlight that bounded subsidies may strictly outperform bounded tolls and outline the proof structure.

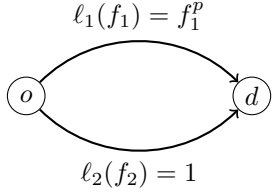
\begin{figure}[t!]
\vspace{2mm}
    \centering
    \begin{tikzpicture}
        \node () [align=center, above] at (1.5,0.75) {$\ell_1(f_1)=f_1^p$};
        \node () [align=center, below] at (1.5,-0.75) {$\ell_2(f_2)= 1$};
        \node (o) [circle,draw,inner sep=0pt,fill=white,minimum width=0.5cm]  at (0.0,0) {$o$};
        \node (d) [circle,draw,inner sep=0pt,fill=white,minimum width=0.5cm]  at (3.0,0.0) {$d$};
        \draw [->,thick,out=45,in=135] (o) edge (d);
        \draw [->,thick,out=-45,in=-135] (o) edge (d);
    \end{tikzpicture}
    \caption{\small Two link parallel congestion game. One edge possesses a polynomial latency function, the other a constant latency function. This routing problem realizes the worst case price of anarchy for polynomial congestion games~\cite{Roughgarden2003}.}
    \label{fig:polypigou_graph}
\end{figure}

\begin{example}\label{ex:poly}
\emph{Polynomial Congestion Network.} Consider a congestion game,  depicted in \cref{fig:polypigou_graph},  possessing two nodes forming a source destination pair with unit mass of traffic and two parallel edges between them, one with latency function $\ell_1(f_1) = f_1^p$, where \BLF{$p$ is a positive integer}, and the other $\ell_2(f_2) = 1$.
This example has been shown to demonstrate the worst case inefficiency among polynomial congestion games~\cite{Roughgarden2003}.

\textit{Step 1: Identify an Optimal Incentive.}
When users are homogeneous in their sensitivity to incentives, an optimal toll for this class of games is the marginal cost toll in \eqref{eq:marginal_cost}, proven to incentivize optimal routing~\cite{Pigou1920}.
Notice that the marginal-cost toll will manifest in this network as 
\begin{eqnarray}
\tau^{\rm mc}_1(f_1) = pf_1^p, & \tau^{\rm mc}_2(f_2) = 0,
\end{eqnarray}
and indeed incentivize the Nash flow to be the system optimal of $f_1 = 1/\sqrt[\leftroot{-3}\uproot{3}p]{p+1}$.

\textit{Step 2: Find Incentives with similar performance.} It can be shown that any incentive mechanism in the set
\begin{equation}\label{eq:set_of_mc}
\left\lbrace T(\ell)= \lambda T^{\rm mc}(\ell) + (\lambda-1)\ell \ | \ \lambda >0 \right\rbrace,
\end{equation}
has the same performance  as the marginal cost taxation mechanism.
This observation can be proven from the later \cref{lem:scaling}.

\textit{Step 3: Identify Bounded Subsidies and Tolls.}
For a bounding factor $\beta \geq p$ the marginal cost taxation mechanism gives a price of anarchy of one; however, for $\beta \in [0,p)$, there exists no taxation mechanism in the set defined in \eqref{eq:set_of_mc} which possesses all optimal mechanisms.
The similar subsidy mechanism
\begin{equation}\label{eq:equivalent_sub}
T^-(\ell) = (1/(p+1) - 1)\ell + (1/(p+1))T^{\rm mc}(\ell),
\end{equation}
which manifests in the network as
\begin{eqnarray}
\tau^{-}_1(f_1) = 0, & \tau^{-}_2(f_2) = \frac{-p}{p+1},
\end{eqnarray}
is in the set of optimal incentive mechanisms and is valid under bounding factors $\beta \geq \frac{p}{p+1}$.
Thus, for bounding factors $\beta \in [\frac{p}{p+1},p)$, there exists a subsidy mechanism that gives price of anarchy one, but there does not exist a tolling mechanism that does the same.
For other bounding factors, the same principles can be followed.
In \cref{subsec:bounded_aff}, the magnitude of the difference of performance between subsidies and tolls is further explored in the context of affine congestion games.
\end{example}

Having concluded \cref{ex:poly}, in \cref{lem:scaling} we show a transformation on incentive mechanisms that does not affect the price of anarchy under homogeneous user sensitivities.
This transformation gives us the important relationship between incentive mechanisms that their performance is not unique and similar performance can be garnered with different magnitudes of transactions.
\begin{lemma}\label{lem:scaling}
Let $T:L(\gee)\rightarrow \mathcal{T}$ be an incentive mechanism over the family of congestion games $\gee$. If another influencing mechanism is defined as $\Tlam(\ell_e) = \lambda T(\ell_e) + (\lambda - 1)\ell_e$ for \BLF{any} $\lambda > 0$, then
\vs \vs
\begin{equation}\label{eq:lem_scale}
\poa(\gee,T) = \poa(\gee,\Tlam).
\end{equation}
\end{lemma}
The proof of \cref{lem:scaling} appears in the appendix.


\vspace{1mm}
\noindent \textit{Proof of \cref{thm:bounded}:}
First, observe that if $\beta=0$ the only permissible incentive function for tolls and subsidies is $\tau_e^+(f_e) = \tau_e^-(f_e) = 0$, i.e., there is no incentive.
Therefore, the left and right hand side of \eqref{eq:thmPoAincentive} equate to the unincentivized case and \BLF{\eqref{eq:thmPoAincentive}} holds with equality.

Let $j_e(f_e) = \ell_e(f_e)+\tau_e(f_e)$ denote the cost a player observes for utilizing an edge $e$ when a mass of $f_e$ users are utilizing it.
The observed cost of a player $x \in N$ can be rewritten as $J_x(P_x,f) = \sum_{e\in P_x} j_e(f_e)$.
In the case where $\beta > 0$, a bounded tolling function on an edge must exist between $\tau_e^+(f_e) \in [0, \beta \cdot \ell_e(f_e)]$, and the edges observed cost satisfies $j_e^+(f_e) \in [\ell_e(f_e), (1+\beta)\cdot \ell_e(f_e)]$.
Similarly, a subsidy function on an edge must exist between $\tau_e^-(f_e) \in [-\beta \cdot \ell(f_e),0]$, and the edges observed cost satisfies $j_e^-(f_e) \in [(1-\beta)\cdot \ell_e(f_e), \ell_e(f_e)]$.

Let $T^+(\ell_e;\beta)$ be a bounded tolling mechanism with edge costs of $j^+_e(f_e)$.
Now, define $\Tlam(\ell_e) = \lambda T^+(\ell_e;\beta) + (\lambda-1) \ell_e$; from \cref{lem:scaling}, $T^+$ and $\Tlam$  have the same price of anarchy for any $\lambda > 0$.
Let $\hat{j}_e$ be the edge cost under influencing mechanism $\Tlam$, from the construction of $\Tlam$
\begin{equation}\label{eq:ftransfer}
\hat{j}_e = \ell_e + \Tlam(\ell_e) = \ell_e + \lambda T^+(\ell_e;\beta) + (\lambda-1) \ell_e = \lambda j_e^+.
\end{equation}

We now look at the cases where $\beta \in (0,1)$ and $\beta \geq 1$ respectively.
When $\beta \in (0,1)$, let $\lambda = (1-\beta)$.
Now, 
\begin{align*}
\hat{j}_e(f_e) = (1-\beta)j_e^+(f_e) &\in [(1-\beta)\ell_e(f_e), (1-\beta^2)\ell_e(f_e)]\\
& \subset [(1-\beta) \ell_e(f_e), \ell_e(f_e)],
\end{align*}
thus the edge costs are sufficiently bounded such that $\Tlam$ is a permissible subsidy mechanism bounded by $\beta$ with the same price of anarchy as $T^+$.
If $\beta \geq 1$ let $\lambda = 1/(1+\beta)$ and get
\begin{align*}
\hat{j}_e(f_e) = \frac{1}{(1+\beta)}j_e^+(f_e) &\in \left[\frac{1}{(1+\beta)}\ell_e(f_e), \ell_e(f_e)\right] \\
& \subset [(1-\beta) \ell_e(f_e), \ell_e(f_e)],
\end{align*}
and again $\Tlam$ is a permissible subsidy mechanism bounded by $\beta$.
By letting $T^+ = T^{\rm opt+}$ we obtain \eqref{eq:thmPoAincentive}.

We have proven that, for $\beta > 0$, if $\poa(\gee,T^{\rm opt-}(\beta)) = \poa(\gee,T^{\rm opt+}(\beta))$, then there exists a $T^{\rm opt-}(\beta)$ that does not achieve the bound.
The contrapositive of this is that if every optimal subsidy achieves the bound, the price of anarchy guarantees are not equal.
In this case, the optimal subsidies are each tightly bounded and $\poa(\gee,T^{\rm opt-}(\beta)) < \poa(\gee,T^{\rm opt+}(\beta))$, proving the final part of~\cref{thm:bounded}.
$\qed$

\subsection{Bounded Incentives in Affine Congestion Games} \label{subsec:bounded_aff}
In Proposition~\ref{prop:aff_bounded}, we explicitly give the price of anarchy bounds of optimal bounded tolls and subsidies in affine congestion games with homogeneous users, again demonstrating the strictly superior performance of subsidies as well as illustrating the magnitude of this difference in performance.
Observe that the optimal subsidy outperforms the optimal toll for each incentive bound, matching the results from~\cref{thm:bounded}.

\label{subsec:aff_bounded}
As a means of illustrating \cref{thm:bounded}, we look at the well studied class of affine congestion games, denoted by \BLF{$$\gee^{\rm aff} := \left\{ G | \ell_e(f_e) = a_e f_e +b_e,a_e,b_e\geq 0,\forall e \in E(G) \right\}.$$}
We include this result to highlight the appreciable gap in performance between subsidies and tolls in this setting.

\begin{figure}[t!]
\vspace{2mm}
    \centering
    \begin{subfigure}[t!]{0.235\textwidth}
        \includegraphics[width=\textwidth]{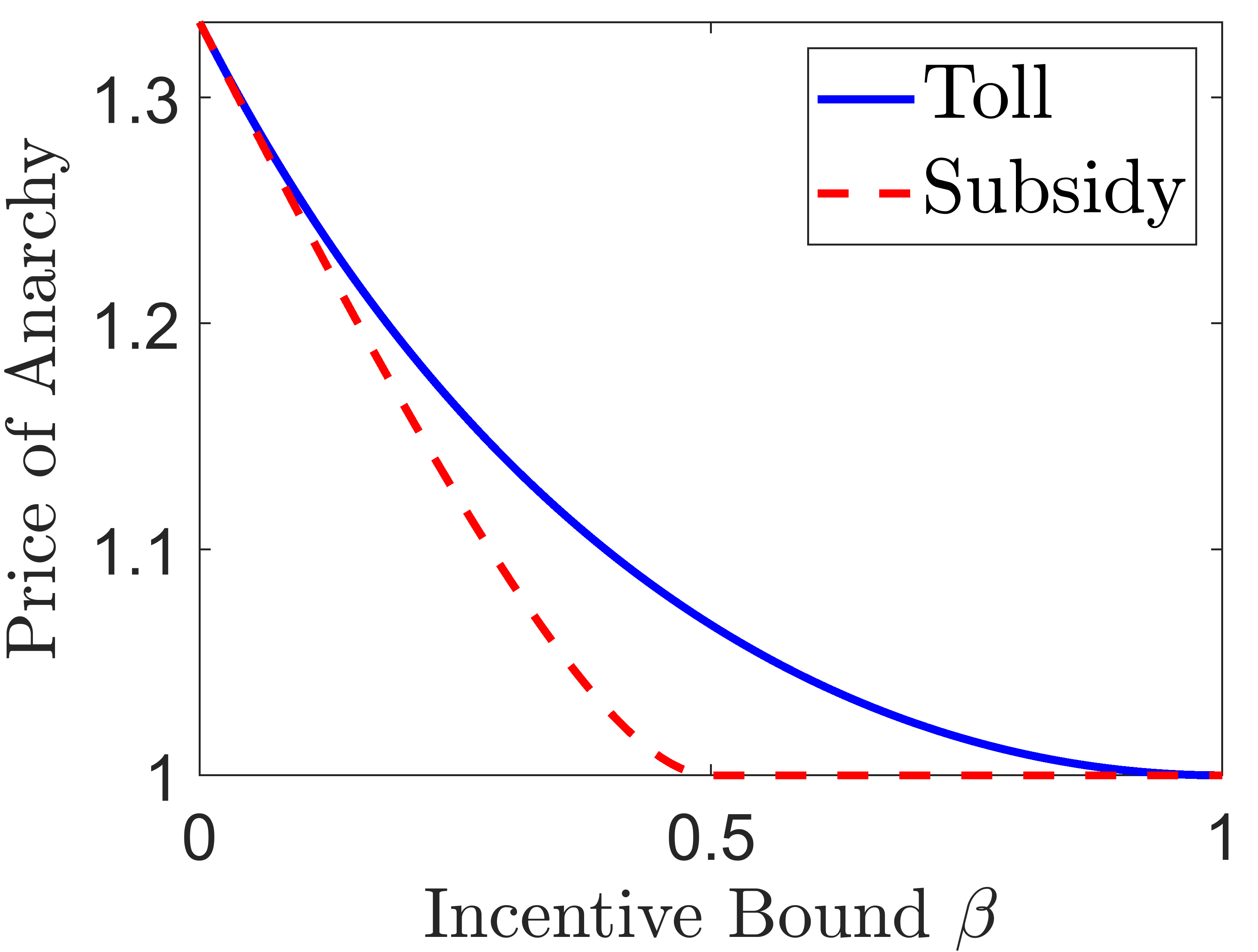}
        \caption{\raggedright \small Price of anarchy with bounded incentives}
        \label{fig:poa_bounded}
    \end{subfigure}
    \begin{subfigure}[t!]{0.235\textwidth}
        \includegraphics[width=\textwidth]{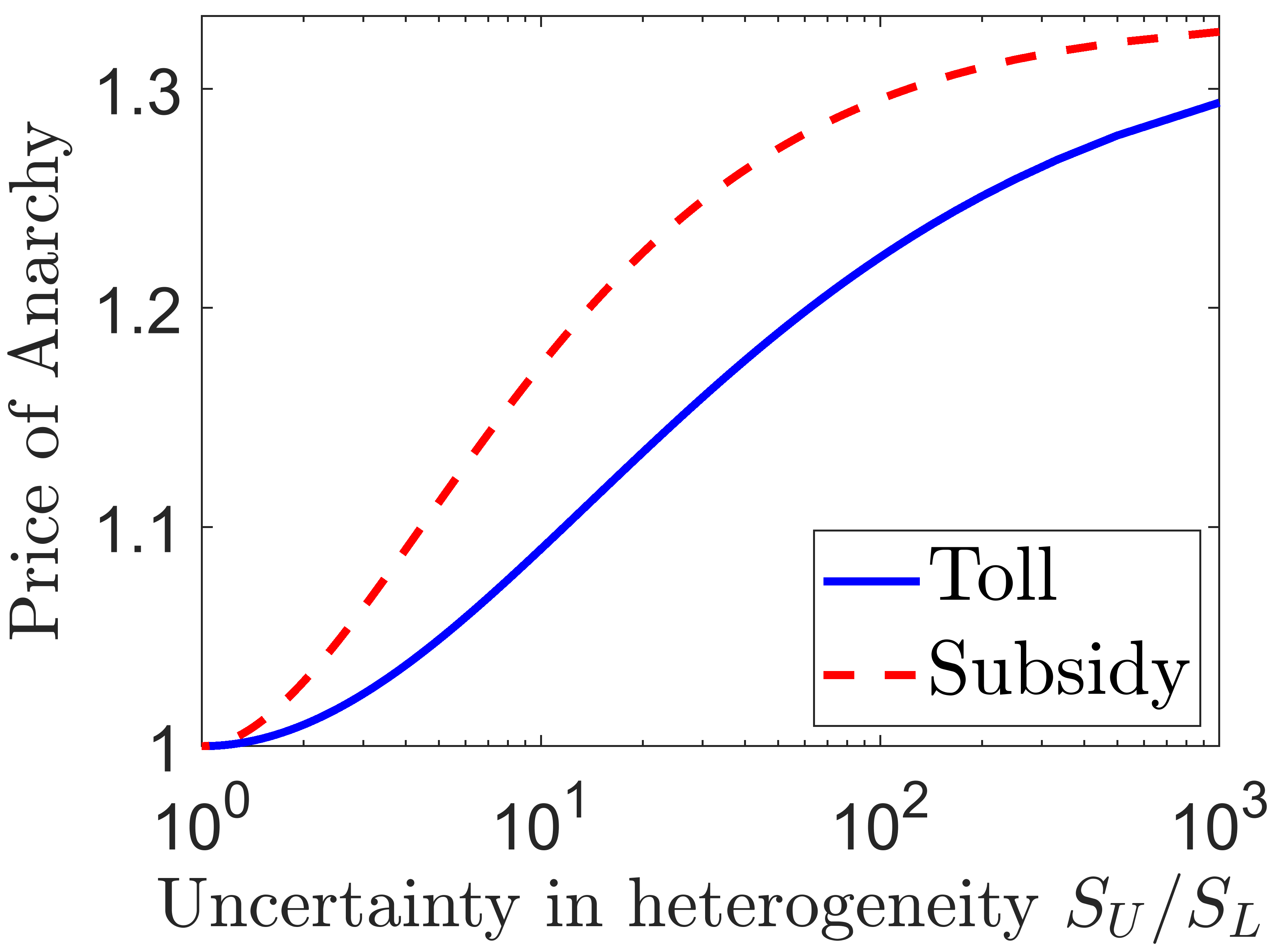}
        \caption{\raggedright \small Price of anarchy with player heterogeneity}
        \label{fig:poa_robust}
    \end{subfigure}
    \caption{\small Price of Anarchy bounds for comparable tolls and subsidies in affine congestion games. (Left) Price of Anarchy under optimal toll and subsidy respectively bounded by a factor $\beta$ from Proposition~\ref{prop:aff_bounded}. (Right) Price of Anarchy of a nominally equivalent toll and subsidy with heterogeneity of user sensitivity introduced from Proposition~\ref{prop:aff_robust}; $\bmax/\bmin$ expresses the amount of possible heterogeneity in the population.}\label{fig:poa}
    \vs \vs \vs \vs \vs \vs
\end{figure}

\begin{proposition}\label{prop:aff_bounded}
	The optimal bounded network-agnostic tolling mechanism in $\gee^{\rm aff}$ is
	\vs \vs
	\begin{equation}
	T^{\rm opt+}(af+b;\beta) = \begin{cases} 
				      \beta a x & \beta \in [0,1), \\
      				  a x & \beta \geq 1, 
   					\end{cases}
	\end{equation}
	with a price of anarchy bound of
	\begin{equation}
	\poa(\gee^{\rm aff},T^{\rm opt+}(\beta)) = \begin{cases} 
				      \frac{4}{3+2\beta-\beta^2} & \beta \in [0,1), \\
      				  1 & \beta \geq 1. 
   					\end{cases}
	\end{equation}
	Additionally, the optimal bounded network-agnostic subsidy mechanism in $\gee^{\rm aff}$ is
	\vs \vs
	\begin{equation}
	T^{\rm opt-}(af+b;\beta) = \begin{cases} 
				      -\beta b & \beta \in [0,1/2), \\
      				  -b/2 & \beta \geq 1/2, 
   					\end{cases}
	\end{equation}
	with a price of anarchy bound of
	\begin{equation}
	\poa(\gee^{\rm aff},T^{\rm opt-}(\beta)) = \begin{cases} 
				      \frac{4}{3+2\hat{\beta}-\hat{\beta}^2} & \beta \in [0,1/2), \\
      				  1 & \beta \geq 1/2, 
   					\end{cases}
	\end{equation}
	where $\hat{\beta} = 1/(1-\beta)-1$. Accordingly, for any $\beta \in (0,1)$,
	\begin{equation}
		\BLF{\poa(\gee^{\rm aff},T^{\rm opt+}(\beta)) > \poa(\gee^{\rm aff},T^{\rm opt-}(\beta)).}
	\end{equation}
\end{proposition}

The proof of Proposition~\ref{prop:aff_bounded} appears in the appendix.
\cref{fig:poa_bounded} illustrates the price of anarchy for tolls and subsidies respectively over various incentive bounds.
Though this result is only for a specific class of games, it helps to quantify the broader notion of \cref{thm:bounded}: \textit{when users are homogeneous in their response to incentives, a subsidy can consistently give price of anarchy closer to one and often by a significant margin.}
In the following sections, we further inspect this relationship when user heterogeneity is introduced.

%% file: Incentives_w_Hetero.tex
\section{Incentives with Heterogeneity}\label{sec:inc_w_hetero}
\cref{sec:bounded} showed that, when users are homogeneous in their response to incentives, subsidies offer better performance guarantees than tolls under budgetary constraints.
We now seek to understand how each type of incentive performs when users differ in their price sensitivity.

Specifically, each user $x \in N$ is associated with a sensitivity $s_x > 0$ to incentives.
We call $s:N \rightarrow \mathbb{R}_{> 0}$ a \textit{sensitivity distribution}. 
We highlight the case where $s_x = c \ \forall x\in N$ for some known constant $c$ as a \textit{homogeneous} distribution of user sensitivities\footnote{\BLF{Without loss of generality, we use $s_x=1$ for a homogeneous population, as was the case in \cref{sec:bounded}}}, in which each user behaves similarly; any other distribution is referred to as a population of \textit{heterogeneous} users.

A user $x \in N_i$ traveling on a path $P_x \in \paths_i$ observes cost
\begin{equation} \label{eq:player_cost_sens}
	J_x(P_x,f) = \sum_{e \in P_x} \ell_e(f_e) + s_x\tau_e(f_e).
\end{equation}
A flow $f$ is a Nash flow if
\begin{multline}
	J_x(P_x,f) \in \argmin_{P \in \paths_i} \left\lbrace \sum_{e\in P} \ell_e(f_e) + s_x\tau_e(f_e) \right\rbrace \\ \forall x\in N_i, ~ i \in \{1,\ldots,k\}.
\end{multline}
A game is now denoted by the tuple $(G,s,\{\tau_e\}_{e \in E})$.

To quantify the robustness of an incentive mechanism, we also consider that the system designer may be unaware of users' response to incentives.
We denote a set of sensitivity distributions by $\sdist = \{s:N \rightarrow [\bmin,\bmax] \}$, where $\bmin>0$ is a lower bound on users' sensitivity to incentives and $\bmax\geq \bmin$ is an upper bound; we include these bounds to quantify the range of users responses, signifying the amount of possible user heterogeneity.

We extend the prior definition of the price of anarchy to include the heterogeneity of users.
Let $\Lnash(G,s,T)$ be the highest total latency in a Nash flow of the game $(G,s,T(L(G)))$.
Now we define,
\vs
\begin{equation}
	\poa(\gee,\sdist,T) = \sup_{G \in \gee} \ \sup_{s \in \sdist} \ \frac{\Lnash(G,s,T)}{\Lopt(G)},
\end{equation}
where the price of anarchy ratio is now the worst case inefficiency over all routing problem, sensitivity distribution pairs using the incentive mechanism $T$.

\BLF{To illustrate this notation, we revisit \cref{ex:homo}, also depicted in \cref{fig:routing_example}, but now with user heterogeneity.

\begin{example}
In the routing problem $G$, depicted in \cref{fig:routing_example}, consider the user sensitivity distribution $s = \{s_x=2 \forall x\in N_1,~ s_x=1/2 \forall x\in N_2\}$.
As a reminder, the optimal flow in $G$ is $\fopt \approx\{ 0.289,0.211,0.25,0.25,0.461 \}$ with a total latency of $\mathcal{L}(\fopt)\approx 0.683$, and with no tolling, the Nash flow is $\fnash=\{ 1/2,0,0,1/2,0 \}$ with total latency $\mathcal{L}(\fnash)=1$ producing a price of anarchy of $\poa(G,s,\emptyset)\approx 1.465$. 
With a marginal cost toll $T^{\rm mc}$ as defined in \eqref{eq:marginal_cost}, the Nash flow becomes $\fnash\approx\{ 0.224,0.276,0.167,0.333,0.443 \}$ producing a price of anarchy of $\poa(G,s,T^{\rm mc})\approx 1.04$. 
With a subsidy mechanism $T^-(\ell_e) = \frac{1}{3}f_e\cdot \frac{d}{df_e}\ell_e(f_e) - \frac{2}{3}\ell_e$ as defined in \eqref{eq:equivalent_sub} with $p=2$, the Nash flow becomes $\fnash\approx\{ 0,0.5,0.137,0.363,0.637 \}$ producing a price of anarchy of $\poa(G,s,T^{\rm sub})\approx 1.32$.
\end{example}

This example shows that user heterogeneity can have a notable impact on the effectiveness of incentives and can effect their relative performance.
In the remainder of this paper, we consider the setting where users are heterogeneous in their price sensitivity when discussing the relative performance of subsidies and tolls.}
We start by looking at tolls and subsidies independently and investigate their performance in the limit of allowable incentives, i.e., as the budgetary constraint is lifted, how does each type of incentive fare?

In \cref{thm:lim} we look at the performance of tolls first and find that, when the budgetary constraint is lifted, tolls can eliminate the negative effect of user heterogeneity.

\begin{theorem}\label{thm:lim}
For a class of congestion games $\gee$, let $T^* \in \arginf_T \poa(\gee,T)$ be an optimal incentive mechanism for homogeneous populations, then
\begin{equation}\label{eq:thm_lim}
	\lim_{\beta \rightarrow \infty} \inf_{T^+ \in \Tset_\beta^+} \poa(\gee,\sdist,T^+) = \poa(\gee,T^*).
\end{equation}
Furthermore, if $\overline{\gee}$ is any class of non-atomic congestion games that has convex, non-decreasing, and continuously differentiable latency functions, then
\begin{equation}
	\lim_{\beta \rightarrow \infty} \inf_{T^+ \in \Tset_\beta^+} \poa(\overline{\gee},\sdist,T^+) = 1.
\end{equation}
\end{theorem}
The proof of \cref{thm:lim} appears in the appendix\footnote{\BLF{This result is reminiscent of \cite{Pigou1920} stating that there exist tolls that influence optimal selfish routing in some settings. In this paper, we extend the result from \cite{Pigou1920} to cases where users are heterogeneous and classes of games where a price of anarchy of one may not be achievable. We note that \cref{thm:lim} is more general than of \cite[Theorem 1]{Brown2017d}, as this result is given for general incentives and is not reliant on marginal cost taxes nor is it limited to the family of congestion games in which they are optimal. Further, the results of \cite{Fleischer2004} cover the case in which the system designer is fully aware of the users' price sensitivities (or value of time in their case) and applies fixed tolls. In contrast, in this paper the toll designer is unaware of the users' exact price sensitivities but is still able to provide a flow-varying tolling scheme that gives a price of anarchy of one as the bounding constraint is lifted.}}.
\BLF{The proof of \cref{thm:lim} follows closely from \cref{lem:hetero} and the notion of responsiveness to heterogeneity presented in the following section.
The result follows from the idea that larger incentives are less impacted by user heterogeneity.
}

After observing positive results for the use of tolls with user heterogeneity, we next seek to understand the effectiveness of subsidies in the same situation.
In \cref{thm:lim_sub}, we show that, even in a restricted class of congestion games, subsidies cannot effectively mitigate the effect of player heterogeneity in the same way tolls do.

\begin{theorem}\label{thm:lim_sub}
Let $\overline{\gee}$ be any class of non-atomic congestion games that has  convex, non-decreasing, and continuously differentiable latency functions, \BLF{the set of latency functions is closed under nonnegative scalar multiplication, and has at least one network where the untolled price of anarchy is greater than one.}
There exists no network-agnostic subsidy mechanism $T$ that gives price of anarchy of 1, i.e.,
\begin{equation}
	\lim_{\beta \rightarrow \infty} \inf_{T^- \in \Tset_\beta^-} \poa(\overline{\gee},\sdist,T^-) > 1.
\end{equation}
\end{theorem}
The proof of \cref{thm:lim_sub} appears in the appendix.

\BLF{Though the class of routing problems has a more strict definition than in \cref{thm:lim}, the result is still very general and holds for most cases other than singleton networks and those where the price of anarchy is always 1.}
From \cref{thm:lim} and \cref{thm:lim_sub} we conclude that \emph{without the presence of budgetary constraints, tolls can mitigate the effect of player heterogeneity while subsidies cannot.}
However, this relationship was shown only as the budgetary constraint was lifted; in the next section, we further investigate the effect of user heterogeneity on subsidies and tolls while budgetary constraints on the incentives remain.

%% file: Robustness.tex
\section{Robustness of Incentives}\label{sec:robust}
In \cref{sec:inc_w_hetero}, user heterogeneity was discussed in the sense of whether incentives could or could not fully mitigate the effect of non-uniform user behavior.
In many cases the very large incentives needed to completely eliminate the negative effects of user heterogeneity are not possible, particularly in the presence of budgetary constraints.
It is thus of interest what the performance guarantees are when the effects of user heterogeneity cannot be entirely overcome and how this compares when using subsidies or tolls.

To compare the robustness of bounded tolls and subsidies, we define an optimal bounded tolling mechanism as 
\begin{equation}
	T^\mathrm{opt+}(\beta,\sdist) \in \underset{T^+ \in \Tset_\beta^+}{\arginf} \ \poa(\gee,\sdist,T^+).
\end{equation}
The optimal bounded subsidy mechanism $T^{\rm opt-}(\beta,\sdist)$ is defined analogously.
For notational convenience, we will omit the dependence on $\sdist$ in the homogeneous setting.

Often, increased user heterogeneity causes performance of an incentive mechanism to diminish. We give the following definition for classes of congestion games with this property.
\begin{definition}\label{def:nonresist}
A class of congestion games is \emph{responsive to player heterogeneity} if $\poa(\gee,\sdist,T^*)$ is strictly increasing with $\bmax/\bmin > 1$ \BLF{for an optimal bounded incentive mechanism $T^* \in \arginf_T \poa(\gee,\sdist,T)$}.
\end{definition}
These classes of games are those that have a degradation in performance from increased player heterogeneity, \BLF{even while the optimal incentive mechanism is in use}; many classes of well studied congestion games possess this property~\cite{Brown2017d}.

\subsection{General Relation of Robustness}
In \cref{thm:robust}, we give a robustness result that shows the performance of subsidies degrades more quickly than tolls as player heterogeneity is introduced.
\begin{theorem}\label{thm:robust}
For a class of congestion games $\gee$, define two incentive bounds $\beta^+$ and $\beta^-$ such that
\vs
\begin{equation}\label{eq:thm_rob_eq}
	 \poa \left(\gee,T^{\rm opt-}(\beta^-)) = \poa(\gee,T^{\rm opt+}(\beta^+) \right),
\end{equation}
\vs
then at the introduction of player heterogeneity,
\begin{equation}\label{eq:thm_rob_uneq}
	\poa \left(\gee,\sdist,T^{\rm opt-}(\beta^-,\sdist)) \geq \poa(\gee,\sdist,T^{\rm opt+}(\beta^+,\sdist) \right) \geq 1.
\end{equation}
Additionally,  each inequality in \eqref{eq:thm_rob_uneq} is strict if $\gee$ is responsive to player heterogeneity and $\bmin < \bmax$.
\end{theorem}
Intuitively, this result stems from the fact that subsidies are more finely tuned to give performance guarantees,
\BLF{as guaranteed in \cref{thm:bounded}. Essentially, applying a small, negative incentive to an edge's increasing latency function will have a more significant impact on the shape of the users' cost function than a larger, positive toll.}
This fact causes the same amount of player heterogeneity to have a larger effect on Nash flows caused by subsidies than with an equivalent toll.
Thus, when increased player heterogeneity escalates the inefficiency, this relationship is strict.
Though the relationship isn't strict for general classes of congestion games, it is for many well studied cases, including the aforementioned polynomial congestion games.

We show in \cref{lem:hetero} a relation between nominally equivalent incentives in the heterogeneous population setting; specifically, we show that the heterogeneous price of anarchy decreases as incentives increase costs to the users.
\begin{lemma}\label{lem:hetero}
For a class of congestion games $\gee$, let $T$ be an incentive mechanism.
If $\Tlam(\ell) = (\lambda-1)\ell + \lambda T$, then $\poa(\gee,\sdist,\Tlam)$ is non-increasing with $\lambda$ and strictly decreasing if $\gee$ is responsive to user heterogeneity and $\bmin < \bmax$.
\end{lemma}
The proof of \cref{lem:hetero} appears in the appendix.

\vspace{1mm}
\noindent \textit{Proof of \cref{thm:robust}:}
First, we give the following definition for incentives that have the same performance in the homogeneous setting.
\begin{definition}
For any incentive mechanism $T$ and $\lambda >0$, each incentive mechanism satisfying $\Tlam(\ell_e) = (\lambda-1)\ell_e + \lambda T(\ell_e)$ is termed \textit{nominally equivalent}.
From~\cref{lem:scaling}, nominally equivalent incentives satisfy 
\begin{equation}
	\poa(\gee,T) = \poa(\gee,\Tlam).
\end{equation}
\end{definition}

The theorem follows closely from \cref{lem:scaling} and \cref{lem:hetero}.
\BLF{First, suppose $T^{\rm opt+}(\beta^+)$ is an optimal tolling mechanism bounded by $\beta^+$.
From \cref{lem:scaling} there exists a nominally equivalent subsidy $\Tlam^-$.
If $\Tlam^- \not\in \Tset_{\beta^-}^-$, then there must exist a $\Tlam^+ \in \Tset_{\beta^+}^+$ that is nominally equivalent to $T^{\rm opt-}(\beta^-)$ from the monotonicity and invertability of the transformation in \cref{lem:scaling}.
From \eqref{eq:thm_rob_eq}, this implies there exists a nominally equivalent $T^{\rm opt+}(\beta^+)$ and $T^{\rm opt-}(\beta^-)$.
}

Now, let $T^{\rm opt-}(\beta^-,\sdist)$ be the optimal subsidy with player heterogeneity bounded by $\beta^-$.
From the fact before, we know there exists a toll $T^+$ that is nominally equivalent to $T^{\rm opt-}(\beta^-,\sdist)$ and bounded by $\beta^+$.
From \cref{lem:hetero}, we obtain that
\vs
\begin{equation}\label{eq:rob_minus}
	\poa(\gee,\sdist,T^+) \leq \poa(\gee,\sdist,T^{\rm opt-}(\beta^-,\sdist)),
\end{equation}
and by the definition of $T^{\rm opt+}(\beta^+,\sdist)$, we get
\vs
\begin{equation}\label{eq:rob_plus}
	\poa(\gee,\sdist,T^{\rm opt+}(\beta^+,\sdist)) \leq \poa(\gee,\sdist,T^+).
\end{equation}
Combining \eqref{eq:rob_minus} and \eqref{eq:rob_plus} gives \eqref{eq:thm_rob_uneq}.
If the class of games is responsive to player heterogeneity, then $\poa(\gee,\sdist,\Tlam)$ is strictly decreasing with $\lambda$ and the relationship is strict.
\hfill $\qed$

\subsection{Robustness of Incentives in Affine Congestion Games}
\cref{thm:robust} states that the performance of subsidies degrades more quickly than tolls when users differ in their response to incentives.
Further, if a subsidy and a toll perform the same in the homogeneous setting, the subsidy performs worse than the toll with any level of user heterogeneity.
To illustrate this fact, we again look at the class of affine congestion games.
In this section, we specifically look at $\gee^{\rm pa}$, defined as the class of parallel-network affine-latency congestion games in which each edge has positive traffic in the untolled Nash flow. We assign taxes using the \BLF{\textit{optimal scaled marginal cost toll with player heterogeneity}}, $T^{\rm smc}(af+b) := (\sqrt{\bmin \bmax})^{-1}af$.
This tolling mechanism was first introduced in~\cite{Brown2017c}, and was shown to minimize the price of anarchy in parallel affine congestion games with sensitivity distributions in $\sdist$ bounded by $\bmin$ and $\bmax$.
In Proposition~\ref{prop:aff_robust}, we give price of anarchy bounds on the optimal scaled marginal cost toll as well as a nominally equivalent subsidy $T^{\rm nes}$.

\begin{proposition} \label{prop:aff_robust}
	Let $\gee^{\rm pa}$ be the set of fully-utilized parallel affine congestion games with sensitivity distributions in $\sdist$. The optimal scaled marginal cost tolling mechanism is \mbox{$T^{\rm smc}(af+b) = \frac{af}{\sqrt{\bmin \bmax}}$} with price of anarchy
	\vs \vs
	\begin{equation}	 \label{eq:poa_smc}
		\poa(\gee^{\rm pa},\sdist,T^{\rm smc}) = \frac{4}{3} \left( 1-\frac{\sqrt{q}}{(1+\sqrt{q})^2} \right).
	\end{equation}
	where $q:=\bmin/\bmax$. Additionally, a nominally equivalent subsidy is $T^{\rm nes}(af+b) = -\frac{1}{1+\sqrt{\bmin \bmax}}b$, with price of anarchy
	\vs \vs
	\begin{equation} \label{eq:poa_nes}
		\poa(\gee^{\rm pa},\sdist,T^{\rm nes}) = \frac{4}{3}\left( 1-\frac{\sqrt{\hat{q}}}{(1+\sqrt{\hat{q}})^2} \right),
	\end{equation}
	where $$\hat{q} = \frac{\lambda q}{1-q+\lambda q} <q,$$
	and $\lambda = \sqrt{\bmin \bmax}/(1+\sqrt{\bmin \bmax})$.
\end{proposition}
The proof of Proposition~\ref{prop:aff_robust} appears in the appendix.
Observe that, because $\hat{q}<q$ in \eqref{eq:poa_smc} and \eqref{eq:poa_nes} the nominally equivalent subsidy has greater price of anarchy when player heterogeneity is introduced.
This can be seen in \cref{fig:poa_robust}.
Intuitively, the same amount of player heterogeneity has a larger effect on the subsidy than the toll.

%% file: Robust_and_bounds.tex
\section{Bounded \& Robust Incentives}\label{sec:bounded_and_robust}
In the previous sections, it was shown that when users are homogeneous in their response to incentives, subsidies offer better performance guarantees than tolls under similar budgetary constraints;
however, as users become heterogeneous in their response to incentives, the performance of subsidies degrades more quickly than that of tolls.
The logical next question we address is, how much heterogeneity causes bounded tolls to outperform bounded subsidies?
In general, this question is difficult to answer.
We therefore look at the case of affine congestion games on parallel networks while using network-agnostic affine incentive functions.
In \cref{thm:robust_and_bounded}, we find the incentive bound $\beta^\ast$ that causes the price of anarchy of the optimal bounded toll and subsidy with user heterogeneity to be equal.
Without loss of generality (because we assume $\bmin$ and $\bmax$ are known to the system designer), we normalize to $\bmin \bmax = 1$.

\begin{figure}[t]
\centering
\includegraphics[width=0.95\linewidth]{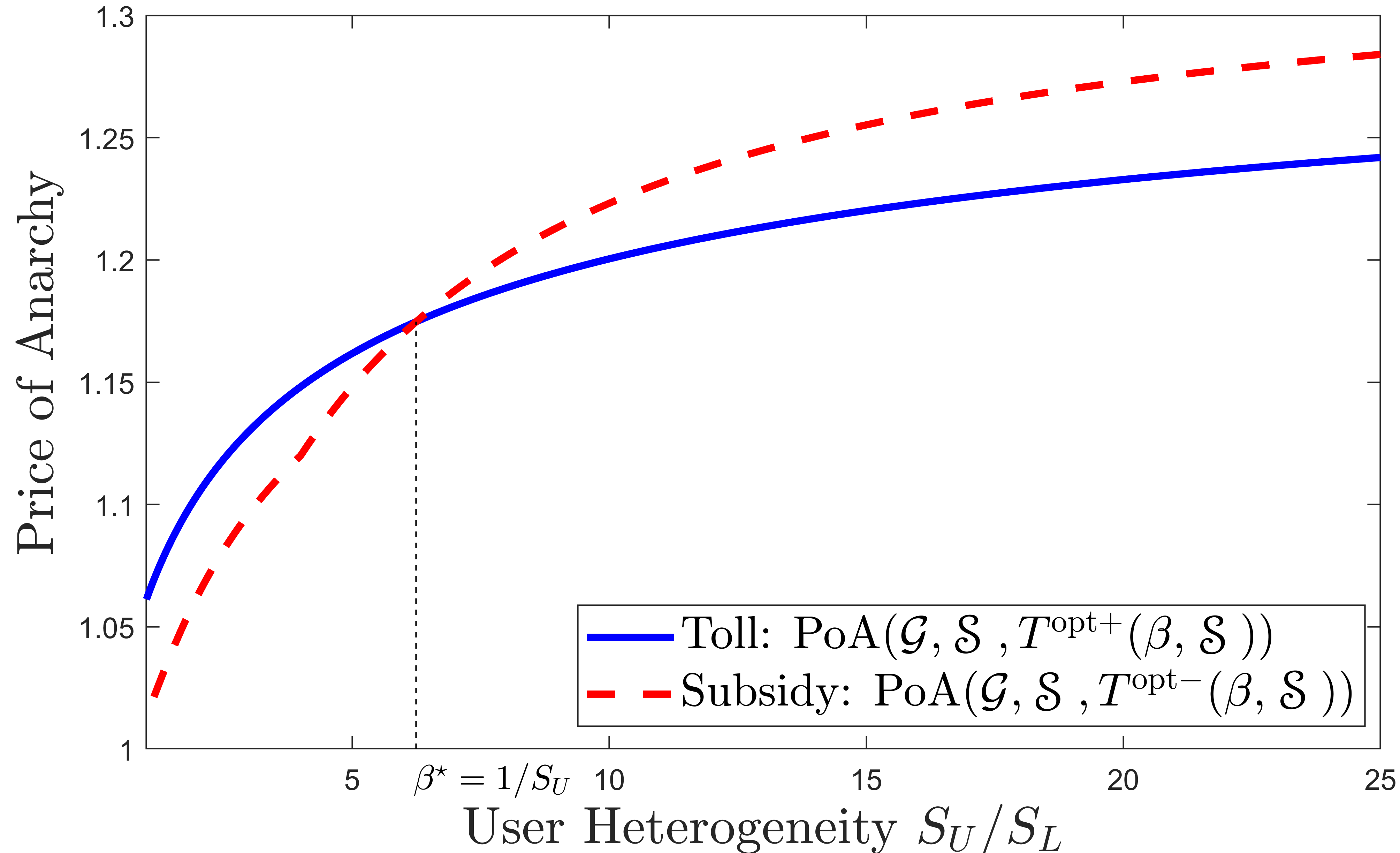}
\caption{\small Price of anarchy under optimal bounded tolls and subsidies with heterogeneous users in parallel-affine congestion games with $\beta = 0.4$. 
When the amount of user heterogeneity is low (i.e. $\bmax/\bmin$ close to one), subsidies offer better performance guarantees than tolls as stated in \cref{thm:bounded}; however, as the level of heterogeneity increases, the performance of subsidies degrade more quickly than tolls, stated in \cref{thm:robust}. 
When the incentive bound \BLF{is} $\beta = 1/\bmax$ the performance of subsidies and tolls is equal, as stated in \cref{thm:robust_and_bounded}.}
\label{fig:opt_robust_bounded}
\end{figure}

\begin{theorem}\label{thm:robust_and_bounded}
Let $T^{\rm opt+}(\beta,\sdist)$ and $T^{\rm opt-}(\beta,\sdist)$ be an optimal, affine toll and subsidy mechanism for $\gee^{\rm pa}$ with incentive bound $\beta$ and player sensitivities between $\bmin$ and $\bmax$.
An incentive bound of \BLF{$\beta^\ast = 1/\bmax = \bmin$} gives
\begin{equation}
\poa(\gee^{\rm pa},\sdist,T^{\rm opt-}(\beta^\ast,\sdist)) = \poa(\gee^{\rm pa},\sdist,T^{\rm opt+}(\beta^\ast,\sdist)).
\end{equation}
\end{theorem}
As illustrated in \cref{fig:opt_robust_bounded}, for lower levels of user heterogeneity (i.e., $\beta^\ast < 1/\bmax$), the optimal subsidy offers price of anarchy closer to one than the optimal toll.
When there is a larger amount of user heterogeneity (i.e., $\beta^\ast > 1/\bmax$) the optimal toll has a lower price of anarchy bound than the optimal subsidy.

The proof of \cref{thm:robust_and_bounded} appears at the end of this section and is supported by the following two propositions.
Proposition \ref{prop:opt_aff_robust_bounded_toll} (originally introduced in \cite{Brown2017d}) gives the optimal affine tolling mechanism and the accompanying price of anarchy guarantee.
\begin{proposition}\label{prop:opt_aff_robust_bounded_toll}
\emph{(Brown \& Marden~\cite{Brown2017d})} Let $T^+(k_1,k_2)$ denote an affine taxation mechanism that assigns tolling functions $\tau_e^+(f_e) = k_1a_ef_e+k_2b_e$. For any $\beta > 0$, the optimal coefficients $k_1^*$ and $k_2^*$ satisfying
\begin{equation}
(k_1^*,k_2^*) \in \argmin_{0 \leq k_1,k_2 \leq \beta} \poa \left( \gee^{\rm pa},\sdist,T^+(k_1,k_2)\right),
\end{equation}
are given by
\begin{align}
	k_1^* &= \beta, \\
	k_2^* &= \max \left\lbrace 0,\frac{\beta^2 \bmin \bmax -1}{\bmin + \bmax + 2 \beta \bmin \bmax} \right\rbrace.
\end{align}
Furthermore, for any $G \in \gee^{\rm pa}$, $\poa(G,\sdist,T^+(k_1^*,k_2^*))$ is upper bounded by the following expression:
\begin{eqnarray}
	\frac{4}{3}\left( 1- \frac{\beta \bmin}{(1+\beta \bmin)^2} \right) & {\rm if} & \beta < \frac{1}{\sqrt{\bmin \bmax}}\label{eq:prop_toll_rb1} \\
	\frac{4}{3}\left( 1- \frac{(1+\beta \bmin)(\frac{\bmin}{\bmax}+\beta \bmin)}{(1+2\beta \bmin + \frac{\bmin}{\bmax})^2} \right) & {\rm if} & \beta \geq \frac{1}{\sqrt{\bmin \bmax}}.\label{eq:prop_toll_rb2}
\end{eqnarray}
\end{proposition}
The proof of Proposition \ref{prop:opt_aff_robust_bounded_toll} appears in the appendix.
The price of anarchy bound is shown in \cref{fig:opt_robust_bounded}.
Similarly, in Proposition \ref{prop:opt_aff_robust_bounded_subsidy} the optimal affine subsidy is given along with its price of anarchy guarantee.
\begin{proposition}\label{prop:opt_aff_robust_bounded_subsidy}
Let $T^-(k_1,k_2)$ denote an affine subsidy mechanism that assigns subsidy functions $\tau_e^-(f_e) = k_1a_ef_e+k_2b_e$. For any $\beta > 0$, the optimal coefficients $k_1^*$ and $k_2^*$ satisfying
\begin{equation}
(k_1^*,k_2^*) \in \argmin_{-\beta \leq k_1,k_2 \leq 0} \poa \left( \gee^{\rm pa},\sdist,T^-(k_1,k_2)\right),
\end{equation}
are given by
\begin{align}
	k_1^* &= 0, \\
	k_2^* &= -\min \left\lbrace \beta, \frac{1}{\bmin + \bmax} \right\rbrace.
\end{align}
Furthermore, for any $G \in \gee^{\rm pa}$, $\poa(G,\sdist,T^-(k_1^*,k_2^*))$ is upper bounded by the following expression:
\begin{eqnarray}
	\frac{4}{3}\left( 1- \beta \bmin(1-\beta \bmin) \right) & {\rm if} & \beta < \frac{1}{\bmin + \bmax}\label{eq:prop_subsidy_rb1} \\
	\frac{4}{3}\left( 1- \frac{\bmin/\bmax}{(1+\bmin/\bmax)^2} \right) & {\rm if} & \beta \geq \frac{1}{\bmin + \bmax}.\label{eq:prop_subsidy_rb2}
\end{eqnarray}
\end{proposition}
The proof of Proposition \ref{prop:opt_aff_robust_bounded_subsidy} appears in the appendix.
The price of anarchy bound is shown in \cref{fig:opt_robust_bounded}.
The price of anarchy bounds equate at $\beta = 1/\bmax$, as substantiated by \cref{thm:robust_and_bounded}, and for $\beta < 1/\bmax$ the subsidy price of anarchy bound is lower, while for $\beta > 1/\bmax$ the toll price of anarchy bound is lower and converging to one.

\vspace{1mm}
\noindent \emph{Proof of \cref{thm:robust_and_bounded}:}
Proposition \ref{prop:opt_aff_robust_bounded_toll} and Proposition \ref{prop:opt_aff_robust_bounded_subsidy} give the price of anarchy bounds for the optimal affine incentives.
By inspection, when $\beta \in [\frac{1}{\bmin + \bmax}, \frac{1}{\sqrt{\bmin \bmax}}]$, the optimal toll and subsidy price of anarchy bounds fall in the domain of \eqref{eq:prop_toll_rb1} and \eqref{eq:prop_subsidy_rb1} respectively.
Additionally, when $\beta = 1/\bmax$, we can see that the optimal toll is $T^+(\frac{1}{\bmax},0)$ and the optimal subsidy is $T^-(0,\frac{-1}{\bmin +\bmax})$; furthermore, these incentives have the same price of anarchy bound, i.e.,
\begin{multline}
\poa\Lp\gee^{\rm pa},\sdist,T^+\Lp\frac{1}{\bmax},0\Rp\Rp \\= \poa\Lp\gee^{\rm pa},\sdist,T^-\Lp 0,\frac{-1}{\bmin +\bmax}\Rp\Rp.
\end{multline}
It is easy to see from \eqref{eq:prop_toll_rb1},\eqref{eq:prop_toll_rb2},\eqref{eq:prop_subsidy_rb1}, and \eqref{eq:prop_subsidy_rb2} that for $\beta > 1/\bmax$,
$$\poa\Lp\gee^{\rm pa},\sdist,T^{\rm opt+}\Rp < \poa\Lp\gee^{\rm pa},\sdist,T^{\rm opt-}\Rp,$$
and for $\beta < 1/\bmax$,
$$\poa\Lp \gee^{\rm pa},\sdist,T^{\rm opt+}\Rp > \poa\Lp \gee^{\rm pa},\sdist,T^{\rm opt-}\Rp.$$
Therefore, $\beta = 1/\bmax$ is the unique incentive bound that gives equal price of anarchy for subsidies and tolls with heterogeneous users in the class of parallel, affine congestion games.
\hfill $\qed$

%% file: Conclusion.tex
\section{Conclusion}
In this work, the effectiveness of subsidies and tolls in congestion games were compared in the presence of budgetary constraints on incentives and user heterogeneity.
The results of this manuscript show that, in a nominal setting, smaller subsidies offer better performance guarantees than tolls; however, in the face of unknown user heterogeneity, tolls are more robust than subsidies.
\BLF{These results hold for general classes of non-atomic congestion games, and future work will show the main conclusions hold for atomic congestion games as well}.
Future work may look at more general notions of user sensitivities as well as other realistic emergent behavior for the society of users.

%% file: Appendix.tex
\appendix
We prove \cref{lem:scaling} using the definition of the Nash flow, and by showing this transformation does not affect user preferences.

\noindent \emph{Proof of \cref{lem:scaling}:}
Let $f^\prime$ be a Nash flow for a game $G \in \gee$ under influencing mechanism $T$.
User $x \in N_i$ observes cost 
\begin{equation}
J_x(P_x,f^\prime) = \sum_{e \in P_x} \ell_e(f^\prime_e) + \tau_e(f^\prime_e),
\end{equation}
and by the definition of Nash flow, will have preferences satisfying
\begin{equation}\label{eq:Nash}
J_x(P_x,f^\prime) \leq J_x(P^\prime,f^\prime), \quad \forall P^\prime\in \paths_i.
\end{equation}
In the same flow $f^\prime$, but now under an influencing mechanism $\hat{T}(\ell_e)= \lambda T(\ell_e) + (\lambda - 1)\ell_e$ where $\lambda>0$, user $x$ observes cost
\begin{align}
\hat{J}_x(P_x,f^\prime) &= \sum_{e \in P_x} \ell_e(f^\prime_e) + \lambda\tau_e(f^\prime_e) + (\lambda-1)\ell_e(f^\prime_e), \\
		&= \sum_{e \in P_x} \lambda(\tau_e(f^\prime_e) + \ell_e(f^\prime_e)) \\
		&= \lambda J_x(P_x,f^\prime).
\end{align}

Observe that through the same process, it can be shown that $\hat{J}_x(P,f^\prime) = \lambda J_x(P,f^\prime)$ for every $P \in \paths_i$.
From \eqref{eq:Nash},
\begin{align}
(1/\lambda)\hat{J}_x(P_x,f^\prime) &\leq (1/\lambda)\hat{J}_x(P^\prime,f^\prime), \quad \forall P^\prime \in \paths_i\\
\hat{J}_x(P_x,f^\prime) &\leq \hat{J}_x(P^\prime,f^\prime), \quad \forall P^\prime \in \paths_i. \label{eq:lamNash}
\end{align}
\eqref{eq:lamNash} holds for all $x\in N$, satisfying that $f^\prime$ is a Nash equilibrium in $G$ under $\hat{T}$.
It is therefore the case that any equilibrium in any game $G \in \gee$ under $T$ is also an equilibrium under $\hat{T}$, thus
\begin{equation}
\Lnash(G,T) = \Lnash(G,\hat{T}),
\end{equation}
and, because this holds for every game $G \in \mathcal{G}$, it certainly holds for the supremum over the set which is the same as \eqref{eq:lem_scale} by definition.
\hfill $\qed$

\noindent \emph{Proof of Proposition \ref{prop:aff_bounded}:}
We first look at the optimal bounded toll and its associated price of anarchy bound.
Trivially, when $\beta > 1$ the optimal toll is the marginal cost toll that gives price of anarchy of one.
For a bounding factor $\beta \in [0,1)$, a feasible bounded toll must satisfy
\begin{equation}
	\tau_e^+(f_e) \in [0,\beta \cdot \ell_e] = [0,\beta a_ef_e+\beta b_e].
\end{equation}
\BLF{Because the tolls are network-agnostic, and must satisfy an additivity property discussed in \cite{Brown2017d} as well as in the proof of \cref{thm:lim_sub}, we can therefore reduce the search for an optimal bounded toll to $\tau_e^+(f_e) = k_1a_ef_e+k_2b_2$ where $k_1,k_2 \in [0,\beta]$.}
We first show that the optimal toll will have $k_2=0$.

Let $T^+$ be a tolling mechanism that assigns bounded tolls with some $k_1,k_2 \in [0,\beta]$.
A player $x \in N_i$ utilizing path $P_x$ in a flow $f$ observes cost
\begin{equation} \label{eq:player_cost_k1k2}
	J_x(P_x,f) = \sum_{e\in P_x}(1+k_1)a_ef_e+(1+k_2)b_e.
\end{equation}
Now, consider an incentive mechanism $\hat{T}$ where edges are assigned tolls $\tau_e(f_e) = (\frac{1+k_1}{1+k_2}-1)a_ef_e$.
Under this new incentive, the same player as before now observes cost
\begin{equation} \label{eq:player_cost_k1}
	\hat{J}_x(P_x,f) = \sum_{e\in P_x}\frac{1+k_1}{1+k_2}a_ef_e+b_e.
\end{equation}
Because the player's cost in \eqref{eq:player_cost_k1k2} and \eqref{eq:player_cost_k1} are proportional, the players preserve the same preferences and the Nash flows remains unaltered.
Because $(\frac{1+k_1}{1+k_2}-1) \leq k_1 \leq \beta$ the new incentive is bounded by $\beta$. 
\BLF{Note that any toll that improves the price of anarchy satisfies $k_1 > k_2$; this can be seen by considering the worst-case example depicted in \cref{fig:polypigou_graph} with $p=1$. Because $0 < k < \beta$,} we need only consider tolls of the form $\tau_e(f_e) = ka_ef_e$ when in search of the optimal bounded toll.
When $k < 0$ the price of anarchy is at least $4/3$ and is indeed not optimal\footnote{Consider the classic Pigou network, as in \cref{fig:polypigou_graph} with $p=1$. It is well known this network gives the worst case price of anarchy of $4/3$ with Nash flow of $f_1=1$. Consider using a taxation mechanism $T(af+b)=kaf$ for some $k<0$ and observe that the Nash flow is unchanged, thus not reducing the price of anarchy for the class of affine congestion games.}.

For a tolling mechanism $T^+(af+b) = kaf$ with $k \in [0,\beta) \subseteq [0,1)$, a player's cost takes the form
\begin{equation}
	J_x(P_x,f) = \sum_{e\in P_x} (1+k) a_e + b_e.
\end{equation}
When player cost functions take this form, the game is similar to that of an altruistic game (introduced in~\cite{Chen2014}) and has price of anarchy of
\vs
\begin{equation}
	\poa(\gee^{\rm aff},T^+) = \frac{4}{3+2k-k^2}.
\end{equation}
The price of anarchy is decreasing with $k \in [0,1)$ and thus the optimal toll occurs when $k$ is maximized at $k=\beta$.

For the optimal subsidy, we now note that incentives must be bounded by $\tau_e(f_e) \in [-\beta \ell_e(f_e),0]$.
From~\cref{lem:scaling}, we can map any such subsidy to an equivalent toll, now constrained to the region $\hat{\tau}_e(f_e) \in [0, \hat{\beta} \ell_e(f_e)]$ where $\hat{\beta} = (\frac{1}{1-\beta}-1)$.
It was shown prior that the optimal tolling mechanism in this region is $\hat{T}(af+b) = \hat{\beta}af$.
Finally, we can again use~\cref{lem:scaling} to map back to the optimal bounded subsidy,
\begin{equation}
	T^{\rm opt-}(af+b) = (\lambda-1)(af+b) + \lambda \hat{T}(af+b),
\end{equation}
with $\lambda = 1-\beta$.
The result is an optimal subsidy of the form $T^{\rm opt-}(af+b)= -\beta b$ for $\beta \in [0, 1/2)$.
The price of anarchy bound comes from considering the equivalent toll.
\hfill $\qed$

\vspace{1mm}
\noindent \emph{Proof of \cref{lem:hetero}:}
First, we assume without loss of generality, that $\bmin=1$.
To see this, we make an equivalent problem where this is true and show the same price of anarchy bound holds.
Let $T$ be any incentive mechanism and $\sdist$ be a family of sensitivity distributions with lower bound $\bmin$ and upper bound $\bmax$.
In any game $G \in \gee$, a player $x \in N_i$ observes costs as expressed in \eqref{eq:player_cost}.
Observe that if we normalize every sensitivity distribution $s \in \sdist$ by multiplying by $1/\bmin$ and correspondingly scale the incentive by $\bmin$ the player cost remains unchanged.
It is therefore the case that any equilibrium is preserved and unchanged, enforcing that
\begin{equation}
	\poa(\gee,\sdist,T) = \poa\left(\gee,\sdist/\bmin,\bmin \cdot T\right).
\end{equation}
Accordingly, we will consider that $\bmin = 1$ throughout.

\BLF{Let $f$ be a flow in $G\in \gee$ induced by sensitivity distribution $s\in \sdist$, and let $T$ be an incentive mechanism that assigns tolls $\tau_e^+$.}
From \cref{lem:scaling} a nominally equivalent incentive mechanism can be found by using the transformation $\hat{T}(\ell_e;\lambda) = (\lambda-1)\ell_e+\lambda T(\ell_e)$, where choosing $\lambda$ sufficiently close to zero causes $\hat{T}$ to be a subsidy mechanism.
We will show that for any $\lambda \in (0,1)$, the incentive mechanism $\hat{T}$ performs worse than $T$ at the introduction of player heterogeneity.

Let $\hat{s}$ be a new sensitivity distribution such that 
\begin{equation} \label{eq:g_sens}
	\hat{s}_x = g(s_x,\lambda)= \frac{s_x}{\lambda + s_x - s_x\lambda},
\end{equation}
\BLF{for all $x \in N$.}
Now, consider an agent's cost in flow $f$ with sensitivity $\hat{s}$ under incentive mechanism $\hat{T}$.
An agent $x \in N_i$ utilizing path $P_x$ in $f$ experiences cost,
\begin{align*}
		\hat{J}_x(P_x,f) &= \sum_{e \in P_x} \ell_e(f_e) + \hat{s}_x\hat{T}(\ell_e(f_e);\lambda)\\
		&= \sum_{e \in P_x} \ell_e(f_e) + \hat{s}_x[(\lambda-1)\ell_e+\lambda \tau_e^+(f_e)]\\
		&= \frac{\lambda}{\lambda + s_x - s_x\lambda} \sum_{e \in P_x}(\ell_e(f_e)+s_x\tau_e(f_e)),
\end{align*}
which is proportional to $J_x(P_x,f)$. By observing proportional costs, players preserve the same preferences over paths, preserving the same Nash flows.

Finally, we show that $\hat{s}$ is a feasible sensitivity distribution in $\sdist$.
From the original bounds $\bmin$ and $\bmax$, any generated distribution $\hat{s}$ exists between $g(\bmin,\lambda)$ and $g(\bmax,\lambda)$.
From before, $\bmin = 1$, thus from \eqref{eq:g_sens}, $g(\bmin=1,\lambda) = 1 = \bmin$, for any $\lambda \in (0,1)$.
Now, observe that any generated distribution satisfies
\begin{equation}
	g(\bmax,\lambda) = \frac{\bmax}{\lambda+\bmax-\bmax \lambda} \leq \bmax,
\end{equation}
for any $\lambda \in (0,1)$.
Thus any generated distribution $\hat{s}$ is sufficiently bounded by $\bmin$ and $\bmax$ and is a feasible distribution in $\sdist$.
By choosing $f$ to be a Nash flow, we can see that any Nash flow that can be induced by some $s \in \sdist$ while using $T$ can similarly be induced by $\hat{s} \in \sdist$ while using $\hat{T}$.
It is therefore the case that the price of anarchy with user heterogeneity is non-decreasing as $\lambda$ decreases, showing the monotonicity.
Further, if $\bmin < \bmax$, then \BLF{$\bmin \leq g(\bmin,\lambda)\leq  g(\bmax,\lambda) < \bmax$,} and if $\gee$ is responsive to user heterogeneity, the price of anarchy is strictly increasing with $\lambda$.
\hfill $\qed$

\vspace{1mm}
\noindent \emph{Proof of \cref{thm:lim}:}
\cref{lem:hetero} states that though two incentive mechanisms have the same price of anarchy when users are homogeneous (from \cref{lem:scaling}), they need not perform the same when users are heterogeneous.
Further, by increasing $\lambda$, one can lower the heterogeneous price of anarchy without altering the performance in the homogeneous setting.
The proof of \cref{thm:lim} is a simple extension of \cref{lem:hetero}.
Increasing $\lambda$ reduces the effect of player heterogeneity on the price of anarchy, and by letting $\lambda \rightarrow \infty$ we can construct an incentive that recovers \eqref{eq:thm_lim}.

In Theorem 1 of \cite{Brown2017d}, the authors propose a realization of this result when using marginal cost taxes.
In the class of congestion games where marginal cost taxes are optimal in the homogeneous setting, they show that the taxation mechanism $$T^u(\ell_e;k)[f_e] = k\left(\ell_e(f_e)+f_e \cdot \frac{d}{df_e}\ell_e(f_e) \right)$$ has a price of anarchy of 1 as $k$ approaches \BLF{infinity}, i.e.,
\begin{equation}
	\lim_{k \rightarrow \infty} \poa(\gee,\sdist,T^u(k)) = 1.
\end{equation}
This same result can be recovered using \cref{thm:lim}.
The marginal cost taxation mechanism defined in \eqref{eq:marginal_cost} has the same performance as $$\Tlam(\ell_e)[f_e] = (\lambda - 1) \ell_e(f_e) + \lambda T^{\rm mc}(\ell_e)[f_e].$$
By taking the limit as \BLF{$\lambda$} approaches infinity, this incentive becomes 
\begin{align*}
\Tlam(\ell_e)[f_e] &= \lambda \left( \ell_e(f_e) + f_e \cdot \frac{d}{df_e}\ell_e(f_e) \right) \\
&= T^u(\ell_e;\lambda)[f_e].
\end{align*}
Not only does this give us the same toll, but by \cref{lem:hetero}, we know that
\begin{equation}
	\lim_{\lambda \rightarrow \infty} \poa(\gee,\sdist,\Tlam) = \poa(\gee, T^{\rm mc}) = 1,
\end{equation}
giving the final statement in \cref{thm:lim}.
\hfill $\qed$

\vspace{1mm}
\noindent \emph{Proof of \cref{thm:lim_sub}:}
First, consider a game $G \in \overline{\gee}$ that has a unique equilibrium and optimal flow respectively, to obtain a heterogeneous price of anarchy of one, the equilibrium must be the same for any sensitivity distribution $s \in \sdist$.
If a taxation mechanism is agnostic of the users' sensitivities, the only way this can be accomplished is by letting the magnitude of the subsidies become large compared to the latency function; for a player $x \in N$ this causes $J_x(P_x,f) \approx \sum_{e \in P_x} s_x T(\ell_e)[f_e]$.
With this subsidy, the users price sensitivity does not affect their preference over paths.

Each of the following three conditions is necessary for a sufficiently large subsidy to incentivize optimal routing (we justify each but note the proof that any one is necessary is trivial).
\begin{enumerate}
\item \emph{Additivity.} A network-agnostic incentive mechanism must satisfy $T(\alpha \ell_1 + \beta \ell_2) = \alpha T(\ell_1) + \beta T(\ell_2)$.
A proof of this appears in \cite{Paccagnan2019I}; 
intuitively, a single latency function can be represented as multiple in series and the total incentive must be the same in both cases to guarantee the same total cost.
\item \emph{Incentives are Unbounded.} $|T(\ell)[f]|>M~\forall~M\in[0,\infty)~\forall \ell \in L(\gee),~f>0$. Any bounded incentive may allow different sensitivity distributions to induce different equilibrium flows.
When the optimal flow is unique, a bounded incentive is incapable of enforcing each equilibrium flow be optimal.
\item \emph{Related by Marginal Cost.} For any two edges $\ell_i,\ell_j$ with respective flow $f_i,f_j$, if $\ell_i^{\rm mc}(f_i) \leq \ell_j^{\rm mc}(f_j)$ then $T(\ell_i)[f_i] \leq T(\ell_j)[f_j]$, where $\ell_i^{\rm mc}(f_i) = \ell_i(f_i) + f_i \frac{d}{df_i}\ell_i(f_i)$ is the marginal cost on edge $i$.
Recall that $T$ is defined as a cost and therefore negative for subsidies, thus this condition states that users must receive less subsidy on edges with higher marginal cost.
It is shown in \cite{Roughgarden2005} that when users observe the marginal cost, the equilibrium flow is optimal.
\end{enumerate}
Note that condition 2 implies player costs are negative: $\ell(f)+T(\ell)[f]<0~\forall \ell \in L(\gee),~f>0$.
Similarly, condition 3 implies that incentives are non-decreasing: if $f_1>f_2$ then $T(\ell)[f_1] \geq T(\ell)[f_2]~\forall \ell \in L(\gee)$.

We now show that no network-agnostic subsidy mechanism can satisfy each of these three conditions.
Assume $T$ is an optimal subsidy mechanism.
By the symmetry of condition 3, we see that if $\ell_i^{\rm mc}(f_i) = \ell_j^{\rm mc}(f_j)$, then $T(\ell_i)[f_i] = T(\ell_j)[f_j]$.
Consider a unit mass of traffic traversing a two link parallel network with edges possessing latency functions $\ell_1$ and $\ell_2$ that are strictly increasing.
Let $f_1$ be the solution to $\ell_1^{\rm mc}(f_1) = \ell_2^{\rm mc}(1-f_1)$, and by condition 3, $T(\ell_1)[f_1] = T(\ell_2)[1-f_1]$.
Now, consider a similar network, but $\ell_2$ is replaced by a scaled latency function $\frac{1}{2} \ell_2$.
Now, define $f_1^\prime$ as the solution to $\ell_1^{\rm mc}(f_1^\prime) = \frac{1}{2}\ell_2^{\rm mc}(1-f_1^\prime)$; from $\ell_1,\ell_2$ strictly increasing, $f_1^\prime< f_1$.
Implied by condition 3, $T(\ell_1)[f_1^\prime] < T(\ell_1)[f_1]$ and $T(\ell_2)[1-f_1] < T(\ell_2)[1-f_1^\prime]$.
From conditions 1 and 2, $T(\ell_2)[1-f_1^\prime] < \frac{1}{2}T(\ell_2)[1-f_1^\prime] = T(\frac{1}{2}\ell_2)[1-f_1^\prime]$.
Put together this gives,
\begin{align*}
	T(\ell_1)[f_1^\prime] < T(\ell_1)[f_1] &= T(\ell_2)[1-f_1] \\
	&< T(\ell_2)[1-f_1^\prime]\\ 
	&< T(\frac{1}{2}\ell_2)[1-f_1^\prime],
\end{align*}
implying $T(\ell_1)[f_1^\prime] \neq T(\frac{1}{2}\ell_2)[1-f_1^\prime]$, contradicting condition 3.
\hfill $\qed$

\vspace{0.1in}
\noindent \emph{Proof of Proposition \ref{prop:aff_robust}:}
The first part of the proposition comes from~\cite{Brown2017c}. We thus find the nominally equivalent subsidy mechanism and find the associated price of anarchy bound.

For notational convenience, let $k = 1/\sqrt{\bmin \bmax}$; the robust marginal cost toll is thus $T^{\rm smc}(af+b) = kaf$.
From \cref{lem:scaling}, we can derive a nominally equivalent subsidy by $T^{\rm nes}(af+b) = (\lambda - 1)(af+b) + \lambda (kaf)$, for any $\lambda > 0$.
By letting $\lambda = 1/(1+k)$, we get the nominally equivalent subsidy to be $T^{\rm nes}(af+b) = -kb/(1+k) = -\frac{1}{1+\sqrt{\bmin \bmax}}b$.

To determine the price of anarchy of $T^{\rm nes}$ with player heterogeneity, we use the result of \cref{thm:robust} to determine the equivalent level of heterogeneity on the nominally equivalent toll, $T^{\rm smc}$.
Let $s \in \sdist$ be a feasible sensitivity distribution, bounded by $\bmin$ and $\bmax$.
As it is defined above, we seek to find the preimage of $[\bmin,\bmax]$ under the function $g(S,1/(1+k))$.
\BLF{Without loss of generality, we normalize $[\bmin,\bmax]$, to $[q,1]$ and look for its preimage.}
Because $g$ is continuous on $S \in [0,1]$, we look at the endpoints of the region.
We first note that $g(1,\lambda) = 1$ for any $\lambda > 0$.
Next, we determine $\hat{q}$ such that $g(\hat{q},\lambda) = q$ as
$$\hat{q} = \frac{\lambda q}{1-q+\lambda q},$$
and by setting $\lambda = 1/(1+k) = \sqrt{\bmin \bmax}/(1+\sqrt{\bmin \bmax})$ recover the equivalent amount of heterogeneity, $\hat{q}$, on $T^{\rm smc}$ as the original subsidy $T^{\rm nes}$ with heterogeneity $q$.
By replacing $q$ with $\hat{q}$ in \eqref{eq:poa_smc} we obtain the price of anarchy for $T^{\rm nes}$ with heterogeneity.
\hfill $\qed$

%

\vspace{0.1in}
\noindent \emph{Proof of Proposition \ref{prop:opt_aff_robust_bounded_subsidy}:}
The proof follows similar steps to that of Proposition \ref{prop:opt_aff_robust_bounded_subsidy}, which appears in \cite{Brown2017d}.
Let $G \in \gee^{\rm pa}$ be a game instance and user be distributed with sensitivity $s \in \sdist$.
Because each network in $\gee$ is parallel, each path constitutes a single edge.
Under an affine subsidy mechanism $T^-(k_1,k_2)$, player $x \in N$ utilizing edge $e$ observes cost
$$J_x(e,f) = (1-k_1s_x)a_ef_e+(1-k_2s_x)b_e,$$
where $k_1,k_2>0$.
Note that scaling users cost functions does not alter their preference over their paths, thus without loss of generality we can write player costs as 
\begin{equation}\label{eq:aff_cost_frac}
J_x(e,f) = \frac{(1-k_1s_x)}{(1-k_2s_x)}a_ef_e+b_e.
\end{equation}

We define a new incentive mechanism $T^\prime(af+b) = k^\prime af$.
Now, let $s^\prime$ be a new sensitivity distribution such that players observe the same cost under $T^\prime$ as they did in \eqref{eq:aff_cost_frac} with sensitivity distribution $s$, i.e.,
\begin{equation}\label{eq:aff_to_mc}
\frac{(1-k_1s_x)}{(1-k_2s_x)} = (1 + k^\prime s_x^\prime).
\end{equation}
The new distribution can be realized by the transformation
\begin{equation}\label{eq:sens_mc_trans}
 s_x^\prime = \frac{s_x(k_2-k_1)}{k^\prime(1-k_2s_x)}.
\end{equation}
The taxation mechanism $T^\prime$ constitutes a scaled marginal cost toll, for which, the following result exists:
\begin{theorem}
\emph{[Brown \& Marden \cite{Brown2017c}]:} For any network $G \in \gee$ with flow on all edges in
an un-tolled Nash flow, and any $s \in \sdist$, any scaled marginal cost taxation mechanism reduces the total latency of any Nash flow when compared to the total latency of any Nash flow associated with the un-tolled case, i.e., for any $k > 0$
\begin{equation}
\Lnash (G,s,T^A(k,0)) < \Lnash (G,s,\emptyset).
\end{equation}
Furthermore, the unique optimal scaled marginal-cost tolling mechanism uses the scale factor
\begin{equation}
	k^\ast = \frac{1}{\sqrt{\bmin \bmax}} = \argmin_{k \geq 0} \{\poa(\gee,\sdist,T^A(k,0))\}.
\end{equation}
Finally, the price of anarchy resulting from the optimal scaled marginal-cost taxation mechanism is
\begin{equation}\label{eq:mc_poa}
\poa(\gee,\sdist,T^A(k^\ast,0)) = \frac{4}{3}\Lp 1- \frac{\sqrt{\bmin/\bmax}}{\Lp 1+\sqrt{\bmin/\bmax} \Rp^2} \Rp \leq \frac{4}{3}.
\end{equation}
\end{theorem}
Because of this, we set $k^\prime = \frac{1}{\sqrt{\bmin^\prime \bmax^\prime}}$ to be the optimal scaled marginal cost taxation mechanism over the new family of sensitivity distributions $\sdist^\prime$, generated by transforming each sensitivity distribution in $\sdist$ as in \eqref{eq:sens_mc_trans}.

Now, in the original subsidy mechanism $T(k_1,k_2)$, let
\begin{equation}\label{eq:k1_from_kp}
	k_1 = k^\prime = \frac{1}{\sqrt{\bmin^\prime \bmax^\prime}}.
\end{equation}
Combining \eqref{eq:sens_mc_trans} and \eqref{eq:k1_from_kp} gives an expression for the accompanying choice of $k_2$ to satisfy \eqref{eq:aff_to_mc},
\begin{equation}\label{eq:k2_of_k1}
k_2 = \frac{k_1^2 \bmin \bmax -1}{ 2k_1\bmin \bmax - \bmin - \bmax}.
\end{equation}

Observe that \eqref{eq:mc_poa} is decreasing with $\bmin / \bmax <1$.
For the similar taxation mechanism $T^\prime$,
\begin{equation}\label{eq:sl_su_prime}
	\frac{\bmin^\prime}{\bmax^\prime} = \frac{\bmin(1-k_2\bmax)}{\bmax(1-k_2\bmin)},
\end{equation}
for $\sdist^\prime$ found by \eqref{eq:sens_mc_trans}.
Notice \eqref{eq:sl_su_prime} is decreasing with $k_2 < \bmax$\footnote{$k_2$ need be less than $\bmax$ for a bounded price of anarchy. A simple construction to show this is Pigou's network in \cref{fig:polypigou_graph} with $s = \bmax$.}.
Therefore, from \eqref{eq:k2_of_k1}, decreasing $k_1$ decreases the price of anarchy; thus picking $k_1^\ast = 0$ is optimal.
Substituting into \eqref{eq:k2_of_k1} gives $k_2^\ast = 1/(\bmin + \bmax)$.
For $\beta < 1/(\bmin + \bmax)$ it is easy to show by a similar transformation that $k_1^\ast = 0$ and $k_2^\ast = \beta$.

Finally, for subsidies of the form $T^A(0,-k)$ with $k < 1$, we show the price of anarchy bound.
Let $s \in \sdist$ be the users' sensitivity distribution.
Let $T^+ = (\lambda -1) \ell(x) + \lambda T^A(0,-k)$ and let $s^\prime$ be a new sensitivity distribution.
Letting $\lambda = 1/(1-k)$ and $s^\prime_x = \frac{s_x(1-k)}{1-k s_x}$,
\begin{align*}
&\ell_e(f_e) + s_x^\prime T^+(\ell_e)[f_e] = \\
& a_ef_e+b_e + \frac{s_x(1-k)}{1-ks_x} \left[ \big( \frac{1}{1-k} - 1 \big) (a_ef_e + b_e) - \frac{k}{1-k} b_e \right] \\
& = \frac{1}{1-ks_x}(a_e f_e + b_e - k s_x b_e) \\
& \propto \ell_e(f_e) + s_x T^A(0,-k)[f_e].
\end{align*}
From player costs being proportional, we can analyze $T^+$ to get the price of anarchy bound.
The new incentive manifests as $T^+(af+b) = \frac{k}{1-k}af$.
Because $k \leq \frac{1}{\bmin + \bmax}$ for an optimal bounded subsidy (and by our assumption that $\bmin \bmax = 1$ for ease of notation),
\begin{equation*}
k \leq \frac{1}{\bmin + \bmax - 1} = \frac{1}{1/\bmax + \bmax -1} \leq 1 = \frac{1}{\sqrt{\bmin \bmax}}.
\end{equation*}
Thus the price of anarchy for $T^+$ is dictated by \eqref{eq:prop_toll_rb1}.
Substituting $\bmin^\prime = \frac{\bmin(1-k)}{1-k \bmin}$ and $\beta = k/(1-k)$ for $k \in [0, \frac{1}{\bmin + \bmax}]$ gives the price of anarchy in Proposition \ref{prop:opt_aff_robust_bounded_subsidy}.
\hfill $\qed$